\documentclass[twocolumn]{aastex61}

\usepackage{amsmath}
\usepackage{graphicx}
\usepackage{natbib}

\begin{document}

    \begin{abstract}
    	We present a study of the dust--to--gas ratios in five nearby galaxies NGC 628 (M74), NGC 6503, NGC 7793, UGC 5139 (Holmberg I),
    	and UGC 4305 (Holmberg II). Using Hubble Space Telescope broad band WFC3/UVIS UV and optical images from the Treasury
    	program LEGUS (Legacy ExtraGalactic UV Survey) combined with archival HST/ACS data, we correct thousands
    	of individual stars for extinction across these five galaxies using an isochrone-matching (reddening-free Q) method. 
    	We generate extinction maps for each galaxy from the 
    	individual stellar extinctions using both adaptive and fixed resolution techniques, and correlate these maps with neutral HI and CO gas maps from literature, including The HI Nearby Galaxy Survey (THINGS) and the HERA
    	CO-Line Extragalactic Survey (HERACLES). We 
    	calculate dust--to--gas ratios and investigate variations in the dust--to--gas ratio with 
    	galaxy metallicity. We find a power law relationship between dust--to--gas ratio and metallicity, consistent with
    	other studies of dust--to--gas ratio compared to metallicity. We find a change in the relation when H$_2$ is not included. This implies that underestimation of
    	$N_{H_2}$ in low-metallicity dwarfs from a too-low CO-to-H$_2$ conversion factor $X_{CO}$ could have produced too low a slope in the derived
    	relationship between dust--to--gas ratio and metallicity. We also compare our extinctions to those derived
    	from fitting the spectral energy distribution (SED) using the Bayesian Extinction and Stellar Tool (BEAST) for NGC 7793 and 
    	find systematically lower extinctions from SED-fitting as compared to isochrone matching.
    
    \end{abstract}
    

    \author{L. Kahre}
    \affiliation{Dept. of Astronomy, New Mexico State University, Las Cruces, NM}
    \email{lkahre@nmsu.edu}
    
    \author{R. A. Walterbos}
    \affiliation{Dept. of Astronomy, New Mexico State University, Las Cruces, NM}
    
    \author{H. Kim} 
    \affiliation{Gemini Observatory, La Serena, Chile}
    
    \author{D. Thilker}
    \affiliation{Johns Hopkins University, Baltimore, MD}
    
    \author{D. Calzetti}
    \affiliation{Dept. of Astronomy, University of Massachusetts-Amherst, Amherst, MA}
    
    \author{J. C. Lee}
    \affiliation{Space Telescope Science Institute, Baltimore, MD}
    
    \author{E. Sabbi}
    \affiliation{Space Telescope Science Institute, Baltimore, MD}
    
	\author{L. Ubeda}
    \affiliation{Space Telescope Science Institute, Baltimore, MD}
    
    \author{A. Aloisi}
	\affiliation{Space Telescope Science Institute, Baltimore, MD}
	
	\author{M. Cignoni}
    \affiliation{Department of Physics, University of Pisa, Largo B. Pontecorvo 3, I56127, Pisa, Italy}
    \affiliation{INFN, Largo B. Pontecorvo 3, I56127, Pisa, Italy}
    \affiliation{INAF -- Osservatorio Astronomico di Bologna, Bologna, Italy}
    
    \author{D. O. Cook}
    \affiliation{California Institute of Technology, Pasadena, CA, USA}
    \affiliation{Department of Physics and Astronomy, University of Wyoming, Laramie, WY, USA}
    
    \author{D. A. Dale}
    \affiliation{Department of Physics and Astronomy, University of Wyoming, Laramie, WY, USA}
    
     \author{B. G. Elmegreen}
    \affiliation{IBM Research Division, T.J. Watson Research Center, Yorktown Heights, NY, USA}
    
    \author{D. M. Elmegreen}
    \affiliation{Department of Physics and Astronomy, Vassar College, Poughkeepsie, NY, USA}
    
    \author{M. Fumagalli}
    \affiliation{Institute for Computational Cosmology and Centre for Extragalactic Astronomy, Department of 
    	Physics, Durham University, Durham, UK}
    	
    \author{J. S. Gallagher III}
    \affiliation{Department of Astronomy, University of WisconsinMadison, Madison, WI, USA}
    
    \author{D. A. Gouliermis}
    \affiliation{Zentrum f\"ur Astronomie der Universit\"at Heidelberg, Institut f\"ur Theoretische Astrophysik, 
    	Albert-Ueberle-Str.\,2, 69120 Heidelberg, Germany}
    \affiliation{Max Planck Institute for Astronomy,  K\"{o}nigstuhl\,17, 69117 Heidelberg, Germany}
    
    \author{K. Grasha}
    \affiliation{Department of Astronomy, University of Massachusetts, Amherst, MA 01003, USA}
    
    \author{E. K Grebel}
    \affiliation{Astronomisches Rechen-Institut, Zentrum f\"ur Astronomie der Universit\"at Heidelberg, 
    	M\"onchhofstr.\ 12--14, 69120 Heidelberg, Germany}
    	
    \author{D. A. Hunter}
    \affiliation{Lowell Observatory, Flagstaff, AZ, US}
    
    \author{E. Sacchi}
    \affiliation{Department of Physics and Astronomy, Bologna University, Bologna, Italy}
    \affiliation{INAF Osservatorio Astronomico di Bologna, Bologna, Italy}
    
    \author{L. J. Smith}
    \affiliation{European Space Agency/Space Telescope Science Institute, Baltimore, MD, USA}
    
    \author{M. Tosi}
    \affiliation{INAF Osservatorio Astronomico di Bologna, Bologna, Italy}
    
    \author{A. Adamo}   
	\affiliation{Department of Astronomy, Oskar Klein Centre, Stockholm University, AlbaNova University Centre, 	
		SE-106 91 Stockholm, Sweden} 
		
	\author{J. E. Andrews}
	\affiliation{Department of Astronomy, University of Arizona, Tucson, AZ, USA}
	
	\author{G. Ashworth}
	\affiliation{Institute for Computational Cosmology and Centre for Extragalactic Astronomy, Department of 
		Physics, Durham University, Durham, UK}
		
	\author{S. N. Bright}
	\affiliation{Space Telescope Science Institute, Baltimore, MD}
	
	\author{T. M. Brown}
	\affiliation{Space Telescope Science Institute, Baltimore, MD}
	
	\author{R. Chandar}
	\affiliation{Department of Physics and Astronomy, University of Toledo, Toledo, OH, USA}
	
	\author{C. Christian}
	\affiliation{Space Telescope Science Institute, Baltimore, MD}
    
    \author{S. E. de Mink}
    \affiliation{Astronomical Institute Anton Pannekoek, Amsterdam University, Amsterdam, The Netherlands}
    
    \author{C. Dobbs}
    \affiliation{School of Physics and Astronomy, University of Exeter, Exeter, UK}
    
    \author{A. S. Evans}
    \affiliation{Department of Astronomy, University of Virginia, Charlottesville, VA, USA}
    \affiliation{National Radio Astronomy Observatory, Charlottesville, VA, USA}
    
    \author{A. Herrero}
    \affiliation{Instituto de Astrofisica de Canarias, La Laguna, Tenerife, Spain}
    \affiliation{Departamento de Astrofisica, Universidad de La Laguna, Tenerife, Spain}
    
    \author{K. E. Johnson}
    \affiliation{Department of Astronomy, University of Virginia, Charlottesville, VA, USA}
    
    \author{R. C. Kennicutt}
    \affiliation{Institute of Astronomy, University of Cambridge, Cambridge, UK}
    \affiliation{Department of Astronomy, University of Arizona, Tucson, AZ, USA}
    
    \author{M. R. Krumholz}
    \affiliation{Research School of Astronomy and Astrophysics, Australian National University, Canberra, ACT 
    	Australia}
    
    \author{M. Messa}
    \affiliation{Department of Astronomy, Oskar Klein Centre, Stockholm University, AlbaNova University Centre, 
    	SE-106 91 Stockholm, Sweden}
    	
    \author{P. Nair}
    \affiliation{Department of Physics and Astronomy, University of Alabama, Tuscaloosa, AL, USA}
    
    \author{A. Nota}
    \affiliation{European Space Agency/Space Telescope Science Institute, Baltimore, MD, USA}
    
    \author{A. Pellerin}
    \affiliation{Department of Physics and Astronomy, State University of New York at Geneseo, Geneseo, NY, USA}
    	
    \author{J. E. Ryon}
    \affiliation{Space Telescope Science Institute, Baltimore, MD}
    
    \author{D. Schaerer}
    \affiliation{Observatoire de Geneve, University of Geneva, Geneva, Switzerland}
    
    \author{F. Shabani}
	\affiliation{Astronomisches Rechen-Institut, Zentrum f\"ur Astronomie der Universit\"at Heidelberg, M
    	\"onchhofstr.\ 12--14, 69120 Heidelberg, Germany}
    
    \author{S. D. Van Dyk}
    \affiliation{PAC/CalTech, Pasadena, CA, USA}
    
    \author{B. C. Whitmore}
    \affiliation{Space Telescope Science Institute, Baltimore, MD}
    
    \author{A. Wofford}
    \affiliation{Instituto de Astronom\'{i}a, Universidad Nacional Aut\'{o}noma de M\'{e}xico, Unidad Acad\'{e}mica 
    	en Ensenada, Km 103 Carr. Tijuana-Ensenada, Ensenada 22860, M\'{e}xico}
    
    \title{Extinction Maps and dust--to--gas Ratios in Nearby Galaxies with LEGUS}

    \section{Introduction}
        \label{sec:intro}
		Dust has large and 
		varied impacts on studies of star formation, chemical evolution, and galaxy evolution. The most obvious of
		these impacts is the absorption and scattering of optical and ultraviolet (UV) light by dust. Typically, up to 50\% of 
		the total stellar energy in a galaxy is attenuated by dust, hampering the interpretation of galaxy spectral
		energy distributions (SEDs) for their fundamental parameters, such as age, stellar population mix, star
		formation rates, and stellar initial mass functions \citep{calzetti01}. This, in turn, impacts our ability 
		to obtain star formation histories and constrain theories of galaxy evolution and star formation. 
		
		Dust also plays a major role in star formation, both as a means of radiative feedback and a catalyst for 
		the formation of molecular hydrogen and other molecules \citep{mathis90, draine03, mckinnon16,
		cazaux09}. Finally, dust and the dust--to--gas ratio have effects on galaxy and chemical evolution models, 
		since dust acts as a sink for many metals \citep{aoyama16, mckinnon16}.
		
		Extinction, as measured towards individual stars, includes effects of absorption 
		and scattering. For an assumed dust population the extinction is linearly 
		proportional to the column density of that population, through the 
		proportionality of optical depth to dust column density. For the Milky Way (MW), 
		\cite{bohlin78} and	\cite{rachford09} give the relation:
	  
		\begin{equation}
			\frac{N_H}{E(B-V)} = 5.8 \times 10^{21} \:\mathrm{H \: cm^{-2} \: mag^{-1}},
	   		\label{eq:dmratiored}
	  	\end{equation}	   

	  	\noindent where $N_H = N_{HI} + 2N_{H_2}$ is the total neutral hydrogen column 
	  	density and $E(B-V)$ is the reddening in the $B-V$ 
	  	color. $N_{HI}$ and $N_{H_2}$ are the column densities of atomic and molecular hydrogen, respectively.
	  	
	  	For a Galactic extinction law ($R_V \mathbf{= \frac{A_V}{E(B-V)}} = 3.1$), equation \ref{eq:dmratiored} gives:
	  
	  	\begin{equation}
			\frac{A_V}{N_H} = \frac{3.1}{5.8 \times 10^{21} \:\mathrm{H \:cm^{-2}\: mag^{-1}}} = 5.3 \times 
				10^{-22} \:\mathrm{mag \: cm^2 \: H^{-1}}
			\label{eq:dmratioext}
	  	\end{equation}
	  
	  	\noindent where $A_V$ is the extinction in the V-band magnitude. Other galaxies with similar metallicities 
	  	are expected to have a similar relation.
	  	
	  	Measurements of the dust--to--gas mass ratio are often done through direct measurements of the mass of 
	  	neutral gas and dust from radio and IR observations, such as in the analysis of \cite{draine07}, 
	  	who give a dust--to--gas mass ratio that varies with oxygen abundance:	
	  	
	  	\begin{equation}
	  		\frac{M_{dust}}{M_H} = 0.010 \frac{(O/H)}{(O/H)_\odot},
	  		\label{eq:dmratiometalsdraine}
	  	\end{equation}
	  	
	  	\noindent where $\frac{M_{dust}}{M_H}$ is the ratio of dust mass to neutral gas mass, and 
	  	$\frac{(O/H)}{(O/H)_\odot}$ is the ratio of the metallicity to solar metallicity.  More recent work suggests the relationship in the MW might be closer to 
	  	$\frac{M_{dust}}{M_H} = 0.0091 \frac{(O/H)}{(O/H)_\odot}$ \citep{draine11,sofiaparvathi09}. 
	  	
	  	Equation \ref{eq:dmratiometalsdraine} is consistent with the observed 
	  	values of the dust--to--gas mass ratio to within a factor of 2 for each galaxy in the \cite{draine07} sample. New 
	  	calibrations by \cite{fanciullo15, planck16adeb} indeed suggest that the \cite{draineli07} models used in 
	  	\cite{draine07} over-predict dust masses by about a factor of 2 regardless of metallicity. 
	  	However, \cite{draine07} had no galaxies with $12+\log{O/H} < 7.6$, and 
	  	\cite{galametz11} and \cite{remy14} found that low metallicity galaxies have a lower-than-expected dust--to--gas
	  	mass ratio. \cite{galametz11} suggest that perhaps the lack of submm maps on some
	  	dwarf galaxies may lead to an underprediction of the dust, while \cite{remy14} suggest that
	  	the decreased dust--to--gas mass ratio is real, and propose a broken power law for the relationship between 
	  	galaxy metallicity and dust--to--gas mass ratio.
	  	
	  	All of the above studies require an assumption of the CO-to-H\textsubscript{2} conversion factor $X_{CO}$,
	  	which also appears to have a dependence on metallicity for non-MW-like environments. \cite{draine07} use
	  	a constant $X_{CO} = 4 \times 10^{20}$ cm\textsuperscript{2} K\textsuperscript{-1} 
	  	(km/s)\textsuperscript{-1}. More recent studies tend to use the lower value of 
	  	$X_{CO} = 2 \times 10^{20}$ cm\textsuperscript{2} K\textsuperscript{-1} (km/s)\textsuperscript{-1} from \cite{leroy09}.
	  	\cite{remy14} compare the \cite{leroy09} fixed $X_{CO}$ with one that varies with 
	  	metallicty using the \cite{schruba12} scaling (O/H)\textsuperscript{-2} for a sample of 126 galaxies over a 2 dex 
	  	metallicity range. This results in a slight decrease in the dust--to--gas mass ratio for a low-metallicity galaxy. \cite{brinchmann13} 
	  	found that using the metallicity-dependent $\log{X_{CO}} = -1.01\log{O/H} + 29.28$ from \cite{boselli02} is required to 
	  	achieve good agreement between dust--to--gas mass ratios obtained from CO and HI and their $X_{CO}$-independent 
	  	method using optical spectroscopy.  
	  	
   	  	The ability to correct photometry for dust is vital for extragalactic studies
   	  	of the ISM and stellar populations. Extinction maps can be used to correct optical (particularly 
   	  	H$\alpha$) and UV images for the effects of dust in order to better determine star formation
   	  	rates and histories in nearby galaxies. This is particularly important for 
   	  	studies of the initial mass function (e.g. \cite{ashworth17}) and mass-to-light ratios, as classifying young massive stars
   	  	often relies heavily on the UV, where extinction is severe.
   	  	
   	  	The ``pair method" is a common way of obtaining reddening corrections for individual stars by 
   	  	comparing a reddened spectrum or observed multi-band photometry to an unreddened spectrum from a star of the same or similar spectral type \citep{elmegreen80, cardelli92}. However, this is difficult for crowded regions, as additional stars will 
   	  	contaminate the spectrum of an individual star \citep{maiz14}. Previous studies, such as the Hubble 
   	  	Tarantula Treasury Project \citep{sabbi13, demarchi16} and the Panchromatic Hubble Andromeda Treasury (PHAT) 	
   	  	survey \citep{dalcanton15} have used NIR photometry and isochrone matching of red clump stars to measure 
   	  	extinctions to circumvent this problem. This is difficult for stars in more distant galaxies, as the lower
   	  	end of the red giant branch is too faint to obtain accurate photometry without extremely deep imaging, and
   	  	crowding of old stars becomes an issue. \cite{zaritsky02} and \cite{zaritsky99} fit 4-band optical photometry of stars
   	  	in the Magellanic Clouds to stellar atmosphere models to obtain extinction corrections and
   	  	generate extinction maps. These studies relied on ground-based photometry, so the SED-fitting method they
   	  	pioneered has not seen much use outside the Local Group until recently. \cite{gordon16} used SED-fitting of 
   	  	the PHAT data to derive extinction corrections for stars in M31, and have very recently made 
   	  	their fitting code publicly available.
   	  	
   	  	Reddening-free Q parameters have been used to identify the spectral type of massive stars and 
   	  	obtain reddening corrections for 
   	  	decades. \cite{johnson53} generated a table of $Q = (U-B) - \frac{E(U-B)}{E(B-V)}(B-V)$ for B- and O-type 
   	  	stars from photometry of nearly 300 stars within the Milky Way. \cite{bergh68} used a similar system to 
   	  	derive reddenings for clusters in the Magellanic Clouds. This early system worked well for B-type stars, 
   	  	however the relation between Q-value and spectral type flattened for O-type stars, making it difficult to distinguish colors and temperatures 
   	  	of the most massive stars. Despite the difficulty in determining spectral types of the most massive stars, 
   	  	reddening corrections are still accurate. More recent work from \cite{kim12} and \cite{blair15} has used isochrone-
   	  	matching and slightly different definitions of Q to more precisely measure intrinsic colors, improving
   	  	upon older tables of spectral types and Q-values.
   	  	
   	  	In this work, we use the photometry of young massive stars described in Section
   	  	\ref{sec:obs} to generate extinction maps for five galaxies using the 
   	  	isochrone-matching method described in \cite{kim12}, and in more detail 
   	  	in Section \ref{sec:extinctcorrs}. In Section \ref{sec:complete}, we discuss the subset
   	  	of stars used to generate the extinction maps shown in Section \ref{sec:extinctmaps}. 
   	  	We calculate dust--to--gas mass ratios for each
   	  	galaxy in Section \ref{sec:gascorrs} and examine the correlation between dust--to--gas mass ratio and 
   	  	metallicity, using metallicity values and neutral gas maps 
   	  	obtained from literature in Section \ref{sec:dgrvmetal}. Finally, we compare
   	  	the extinctions obtained from isochrone-matching to those obtained via the
   	  	SED-fitting method used in the Bayesian Extinction And Stellar Tool (BEAST) in
   	  	Section \ref{sec:beastcomp}. 
        
    \section{Observations and Galaxy Sample}
        \label{sec:obs}
        The Legacy ExtraGalactic Ultraviolet Survey (LEGUS) has observed 50  
        star-forming galaxies at distances ranging from approximately 3 to 
        16 Mpc. For each galaxy, LEGUS has generated 5-band stellar and cluster photometric 
        catalogs with a combination of the Wide-Field Camera 3 (WFC3) and Advanced Camera 
        for Surveys (ACS) instruments \citep{adamo17, sabbi17}. Our 
        wavelength coverage runs from the near-ultraviolet to the I band, using the 
        F275W (NUV), F336W (U), F438W (B), F555W (V), and F814W (I) filters for the WFC3 
        instrument and the F435W, F555W and F814W filters for the ACS instrument. The 
        survey is described in full in \cite{calzetti15}. We achieve greater
        resolution with HST imaging compared to ground-based observations, providing us 
        with the ability to obtain individual stellar photometry for the most nearby galaxies. 
        
        In general, the LEGUS observations were designed
        to reach a depth of $m_{F275W} = 26.0$ mag with signal-to-noise=6, with similar
        depth in the other filters. Individual stellar positions and fluxes were 
        determined  using the ACS and WFC3 PSF-fitting modules provided in the DOLPHOT 
        photometric package \citep{dolphin02}. Sources with a robust detection in a band do not deviate
        more than $3\sigma$ from the average $\chi^2$ value for the PSF fit for the 
        entire galaxy and have a photometric error of less than 0.5 magnitudes \citep{sabbi17}. 
        The version of the LEGUS stellar catalogs used in this analysis is described
        in \cite{sabbi17}. We use the V2 catalogs, which contain only the stars with the highest
        quality photometry, and we limit our sample to stars with robust detections in the
        NUV, B, V, and I-band filters. These catalogs are uncorrected for foreground MW or local extinction. Both
        of these corrections are discussed in Section \ref{sec:extinctcorrs}. Table
        \ref{tab:hstimages} lists the pointings used in this study and the instrument and 
        filter combinations used for each pointing.
        
        \begin{deluxetable}{m{1.75cm}m{1.2cm}m{1.2cm}m{1.2cm}m{1.2cm}m{1.2cm}}
        \tablecolumns{6}
        \tablecaption{HST LEGUS Filters\label{tab:hstimages}}
        \tablehead{
        	\colhead{Pointing} & \colhead{NUV} & \colhead{U} & \colhead{B}
        	& \colhead{V} & \colhead{I} \\
        }
        \startdata
        \centering NGC 628 (Central) & \centering WFC3 F275W & \centering WFC3 F336W 
        	& \centering ACS F435W & \centering ACS F555W & ACS F814W \\
        \centering NGC 628 (Eastern) & \centering WFC3 F275W & \centering WFC3 F336W 
        	& \centering ACS F435W & \centering WFC3 F555W & ACS F814W \\
        \centering NGC 6503 & \centering WFC3 F275W & \centering WFC3 F336W 
        	& \centering WFC3 F438W & \centering WFC3 F555W & WFC3 F814W \\
        \centering NGC 7793 (Eastern) & \centering WFC3 F275W & \centering WFC3 F336W 
        	& \centering WFC3 F438W & \centering WFC3 F555W & WFC3 F814W \\
        \centering NGC 7793 (Western) & \centering WFC3 F275W & \centering WFC3 F336W 
        	& \centering WFC3 F438W & \centering ACS F555W & ACS F814W \\
        \centering UGC 4305 (Ho II) & \centering WFC3 F275W & \centering WFC3 F336W 
        	& \centering WFC3 F438W & \centering ACS F555W & ACS F814W \\
		\centering UGC 5139 (Ho I) & \centering WFC3 F275W & \centering WFC3 F336W 
			& \centering WFC3 F438W & \centering ACS F555W & ACS F814W \\
		\enddata
		\end{deluxetable}
        
        In this work, we focus on 5 of the 50 galaxies in the LEGUS sample: NGC 628, 
        UGC 4305 and UGC 5139 (Holmberg II and Holmberg I), NGC 6503, and NGC 7793. 
        Table \ref{tab:galprops} lists the properties of our galaxy sample. The five 
        galaxies together represent a
        good subset of the LEGUS sample in distance, inclination, metallicity and 
        size. NGC 628, NGC 6503, and NGC 7793 are all spiral galaxies with a range of masses, 
        though NGC 7793 is categorized as a flocculent. UGC 4305 and UGC 5139 are dwarfs. 
        The three spirals are at a range of distances with NGC 6503 roughly at the middle
        of the LEGUS distance range. NGC 628 is one of the most distant galaxies in the sample, 
        and NGC 7793 is relatively nearby, at roughly
        at the same distance as the two dwarfs. NGC 6503 is also the most inclined of the 
        LEGUS galaxies at about 70 degrees, while the other four can be approximated as 
        face-on. NGC 628 and NGC 7793's large sizes ($\sim$12' and $\sim$14') and multiple LEGUS
        pointings provide us 
        with an opportunity to study radial variations within each galaxy. Both galaxies 
        also have published metallicity gradients from \cite{sanchez13} and \cite{stanghellini15}, 
        providing us a larger sample of metallicities for 
        this study. Later studies will expand upon this sample.

        \begin{deluxetable*}{ccccccccccccccc}
        	\tablewidth{\textwidth}
        	\tablecolumns{12}
        	\tablecaption{Properties of the LEGUS Galaxy Sample. Reproduced from \cite{calzetti15}\label{tab:galprops}}
			\tablehead{
				\colhead{Name\tablenotemark{a}} & \colhead{v\textsubscript{H}\tablenotemark{a}} 
				& \colhead{Morph.\tablenotemark{a}} & \colhead{T\tablenotemark{b}}
				& \colhead{Inclin.\tablenotemark{a}} & \colhead{Dist.\tablenotemark{c}} 
				& \colhead{Method\tablenotemark{d}} 
				& \multicolumn{2}{c}{12+log(O/H)\tablenotemark{e}} 
				& \colhead{SFR(UV)\tablenotemark{f}} 
				& \colhead{M\textsubscript{*}\tablenotemark{g}} 
				& \colhead{M(HI)\tablenotemark{h}} \\
				\colhead{} & \colhead{(km s\textsuperscript{-1})} & \colhead{} & \colhead{}
				& \colhead{(degrees)} & \colhead{(Mpc)} & \colhead{} & \colhead{(PT)} & \colhead{(KK)}
				& \colhead{(M\textsubscript{$\odot$} year\textsuperscript{-1})} 
				& \colhead{(M\textsubscript{$\odot$})} & \colhead{(M\textsubscript{$\odot$})}}%\\
%				\colhead{(1)} & \colhead{(2)} & \colhead{(3)} & \colhead{(4)} & \colhead{(5)} 
%				& \colhead{(6)} & \colhead{(7)} & \colhead{(8)} & \colhead{(9)} & \colhead{(10)} 
%				& \colhead{(11)} & \colhead{(12)}}
			\startdata
			NGC 0628 & 657 & SAc & 5.2(0.5) & 25.2 & 9.9 & SNII & 8.35 & 9.02 & 3.67 & 1.1E10 & 1.1E10 \\
			NGC 6503 & 25 & SAcd & 5.8(0.5) & 70.2 & 5.27 & TRGB & \multicolumn{2}{c}{8.69} & 0.32 & 1.9E09 & 1.3E09 \\
			NGC 7793 & 230 & SAd & 7.4(0.6) & 47.4 & 3.44 & Ceph & 8.31 & 8.88 & 0.52 & 3.2E09 & 7.8E08 \\ 		
			UGC 4305 & 142 & Im & 9.9(0.5) & 37.1 & 3.05 & Ceph & \multicolumn{2}{c}{7.92} & 0.12 & 2.3E08 & 7.3E08 
				\\
			UGC 5139 & 139 & IABm & 9.9(0.3) & 33.6 & 3.98 & TRGB & \multicolumn{2}{c}{8.00} & 0.02 & 2.5E07 & 	
				2.1E08 \\
			\enddata

		\tablenotetext{a}{Galaxy name, recession velocity, and morphological type as listed in NED, the NASA Extragalactic 				Database. Inclination, in degrees, derived from the sizes listed in NED.}
		\tablenotetext{b}{RC3 morphological T-type as listed in Hyperleda (http://leda.univ-lyon1.fr) and discussed in 		
			\cite{kennicutt08} for the Local Volume Legacy Survey (LVL) galaxies, from which the LEGUS sample is derived. In that paper, T-type 	
			= 11 is adopted for galaxies misclassified as early types while being compact irregular or Blue Compact 
			Galaxies (BCGs). Uncertainties on the morphological classification are in parenthesis. Some of the 
			galaxies have large uncertainties, and they may be misclassified.}
		\tablenotetext{c}{Redshift-independent distance in megaparsecs.}
		\tablenotetext{d}{Methods employed to determine the distances. In order of decreasing preference: Cepheids (Ceph), 
			Tip of the Red Giant Branch Stars (TRGB), Surface Brightness Fluctuations (SBF), and Supernova Type II 	
			Plateau (SNII).}
		\tablenotetext{e}{Characteristic oxygen abundances of the galaxies. For NGC 628 and NGC 7793, this is the globally averaged 
			abundance (\cite{moustakas10}). The two columns, (PT) and (KK), are the oxygen abundances on two 
			calibration scales: the PT value, in the left-hand-side column, is from the empirical calibration of 
			\cite{pilyugin05}, the KK value, in the right-hand-side column, is from the theoretical 
			calibration of \cite{kobulnicky04}. The dwarves and NGC 6503 have only one metallicity available (via 
			the ``direct" method, from \cite{croxall09} and \cite{tikhonov14}), so the value straddles the two 
			columns.}
		\tablenotetext{f}{Star formation rate (M$_\odot$ year\textsuperscript{-1}), calculated from the GALEX far-UV, corrected for 
			dust attenuation, as described in \cite{lee09}.}
		\tablenotetext{g}{Stellar masses (M$_\odot$), obtained from the extinction-corrected B-band luminosity, and color 	
			information, using the method described in \cite{bothwell09} and based on the mass-to-light ratio 	
			models of \cite{bell01}.}
		\tablenotetext{h}{HI masses, using the line fluxes listed in NED, applying the standard formula: $M(HI)[M_\odot] = 2.356 \mathbf{\times} 
			10^5 D^2 S$, where D is the distance in megaparsecs, and S in the integrated 21 cm line flux in units of 	
			Jy cm s\textsuperscript{-1}.}
	\end{deluxetable*}  
        
    \section{Extinction Corrections}
        \label{sec:extinctcorrs}
        Since the LEGUS stellar photometric catalogs are not corrected for the foreground MW extinction, we first 
        find the foreground MW extinction from the values in \cite{schlafly11}, converted to the HST filters for 
        each galaxy. We then apply those extinctions to the stars in each galaxy's photometric catalog.
        Next, using the method described in \cite{kim12}, we match the 5-band photometry to theoretical isochrones
        to obtain extinction corrections for individual stars. First, we calculate
        reddening-free Q values for massive stars using equation \ref{eq:qvalue}:

          \begin{equation}
                Q_{NBVI} = (\mathbf{NUV}-B) - \frac{E(\mathbf{NUV}-B)}{E(V-I)}(V-I)
            \label{eq:qvalue}
          \end{equation}
      
          \noindent where NUV, B, V, and I are the observed magnitudes in the HST filter 
          corresponding to that band, and E(NUV-B) and E(V-I) are the selective extinctions 
          in those colors. The selective extinction ratio is obtained using a standard 
          MW extinction law ($R_V = 3.1$), which corresponds to $\frac{E(\mathbf{NUV}-B)}{E(V-I)} = 
          1.5999$ \citep{cardelli89}. Table \ref{tab:extlaw} shows
          the extinction coefficients for the 5 bands in the LEGUS
          photometry.  Changes in the choice of the extinction law effectively change the slope 
          of the reddening vector (i.e., slopes of constant Q lines), which can cause changes in the 
          derived extinction by a factor of 2 or more in the most extreme cases. We test
          the effects of this shift in Section \ref{sec:gascorrs} on our extinction maps, finding that the change
          in the derived extinction for the full map is negligible, as the extreme stars are averaged out. We calculate
          the selective extinction ratio and Q based on the center wavelength of each bandpass.
          There is a slight variation of Q if it is calculated based on coefficients calculated by
          modeling the shape of the spectrum in each bandpass, however this effect is negligible compared to 
          the effect of changing the extinction law.
          
          The reddening-free Q value is defined so that it only depends on the spectral type
          of a star, and can be defined differently depending on the availability of photometric
          bands. \cite{binney98} use the UBV system in their definition, while \cite{kim12} use 
          a similar Q to equation \ref{eq:qvalue}, but substitute the U-band for our NUV-band,
          which has higher signal-to-noise ratio in the dataset used in their work than 
          in the NUV-band. In the LEGUS U-band photometry, we also find lower photometric 
          errors than we see in the NUV-band, with average errors approximately 0.02 less in the U-band. 
          However, we use the NUV-band because NUV is more sensitive to extinction, 
          allowing us to measure even small amounts of dust. Massive stars also have the peak of their 
          light distribution further in the UV, making it easier for us to differentiate masses. For stars in 
          the range $10-30 M_\odot$, the Q-value employed by \cite{kim12} varies by approximately 0.2, whereas
          the Q-value defined in equation \ref{eq:qvalue} varies by nearly 0.3 over the same mass range.
          
          \begin{deluxetable}{ccc}
			\tablewidth{\textwidth}
			\tablecolumns{3}
			\tablecaption{Extinction Coefficients\label{tab:extlaw}}
			\tablehead{
				\colhead{HST Filter} & \colhead{$\mathbf{\lambda_{central} (\AA)}$\tablenotemark{a}} & \colhead{$\frac{A_\lambda}{A_V}$\tablenotemark{b} }
			}
			\startdata
				F275W (NUV) 						& 2750	& 1.9893 \\
				F336W (U) 							& 3360	& 1.6328 \\
				F438W/F435W (B)\tablenotemark{c} 	& 4380	& 1.3319 \\
				F555W (V)\tablenotemark{d} 			& 5550	& 0.9886 \\
				F814W (I)\tablenotemark{d} 			& 8140	& 0.5777 \\
			\enddata
			\tablenotetext{a}{Approximate central wavelength of the filter in angstroms.}		
			\tablenotetext{b}{$\frac{A_\lambda}{A_V}$ calculated using the approximate center wavelength
			of the filter.}
			\tablenotetext{c}{The F435W filter is the ACS filter corresponding to the 
			same approximate wavelength range as the WFC3 F438W filter.}
			\tablenotetext{d}{The F555W and F814W filters share the same name and
			wavelength range on both the ACS and WFC3 instruments.}
		  \end{deluxetable}
      
\begin{figure}[h!]
          \centering
            \includegraphics[width=0.5\textwidth]{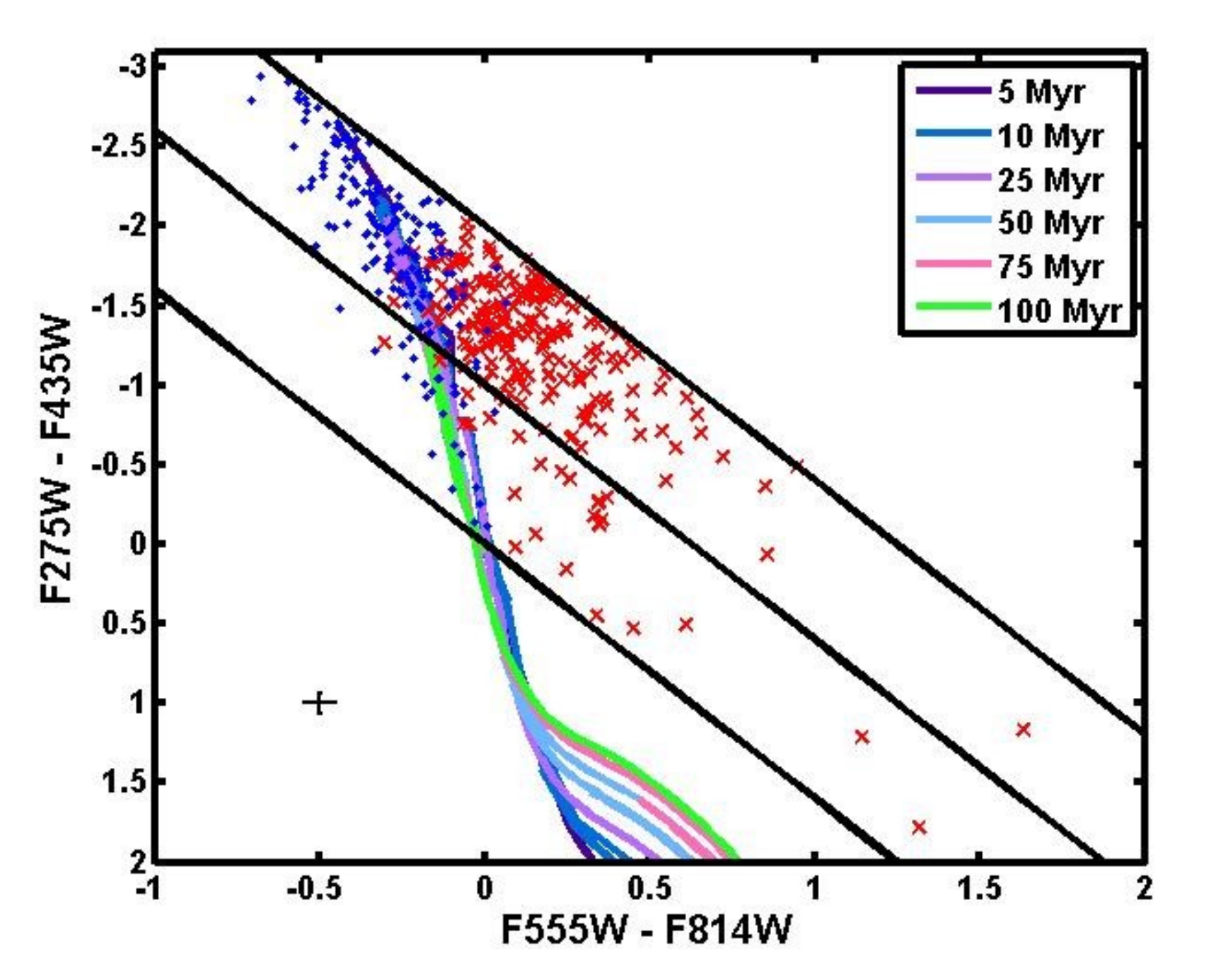}
            \caption{Color-color diagram of 200 stars in NGC 628. Red crosses are corrected for MW extinction, but 
            uncorrected for internal extinction. Blue diamonds are corrected for MW and internal extinction. The black 
            errorbars to the bottom left show the typical photometric
            error from LEGUS. The black diagonal lines are lines of constant Q at 0, -1.0, and -2.0 (starting
            from the lower left and moving up and right), determined by
            equation \ref{eq:qvalue}. The lowest and highest lines show the limits of our extinction correction.
            The colored lines are PARSEC isochrones of various ages with solar metallicity ($Z = 0.0152$) \cite{parsec}.}
            \label{fig:ccdcorr}
          \end{figure}

          Once Q is obtained, we then determine the extinction values for each star by matching each
          individual star's Q-value to the Q-value corresponding to the theoretical isochrones. For stars
          more massive than $2 M_\odot$, the isochrones for the various luminosity classes are very
          similar, therefore we do not initially need to know the luminosity class of the star in order
          to determine the star's intrinsic color. This region is enclosed by the black solid lines in Figure
          \ref{fig:ccdcorr}. To limit the analysis to these stars only, a cut of $Q \leq
          0$ is applied. We also have a limit on Q in the high-mass end of $Q \geq -2.1$, as this is the lowest
          Q for which the isochrones have a mass $\leq 100 M_{\odot}$. \cite{kim12} uses a shallower reddening
          slope for these stars to attempt to find a physical Q-value, however since this should not affect
          the derived extinctions in the maps from section \ref{sec:extinctmaps} significantly, we exclude them for 
          the sake of simplicity. We simultaneously match stars to the 1, 2, 5, 10, 15, 25, 50, 75, and 100 Myr
          isochrones, choosing the isochrone that contains the Q-value closest to that of each individual star.
          
          We can then obtain the intrinsic color directly from the matched isochrone, thus the color 
          excess for each star becomes:
          
          \begin{equation}
              E(V-I) = (V-I)_{observed} - (V-I)_{intrinsic}
              \label{eq:vicolorexcess}
          \end{equation}          
          
          \noindent and similarly for the $(\mathbf{NUV}-B)$ color. We then calculate the extinction values for each
          wavelength band using a standard MW extinction law (\cite{cardelli89}) and obtain corrected magnitudes
          for each star. Stars that are located to the left of the isochrones before the correction are given 
          $E(V-I) = 0$ if they are within their photometric error of the isochrones, otherwise they are removed
          from the corrected catalog. Figure \ref{fig:ccdcorr} shows a sample extinction correction for 200 stars 
          in NGC 628. The extinction correction for each star can be visualized as shifting the star along its 
          constant $Q_{NBVI}$ line to the locus of isochrones in the two-color plot; the distance of the shift is
          proportional to the reddening in each of the two colors on the axes. 
          
          Note that the corrected stars in Figure \ref{fig:ccdcorr} do not lie precisely on the isochrones.
          The stars in the LEGUS catalog each have some level of photometric uncertainty, which we must take into
          account in our correction. After matching a star to an isochrone, we apply a shift in color based on the 
          photometric uncertainty in the four bands used for correction. The final colors and extinctions for 
          corrected stars reflect the correction plus the photometric uncertainty of each star.

\begin{figure}[h!]
          \centering
            \includegraphics[width=0.5\textwidth]{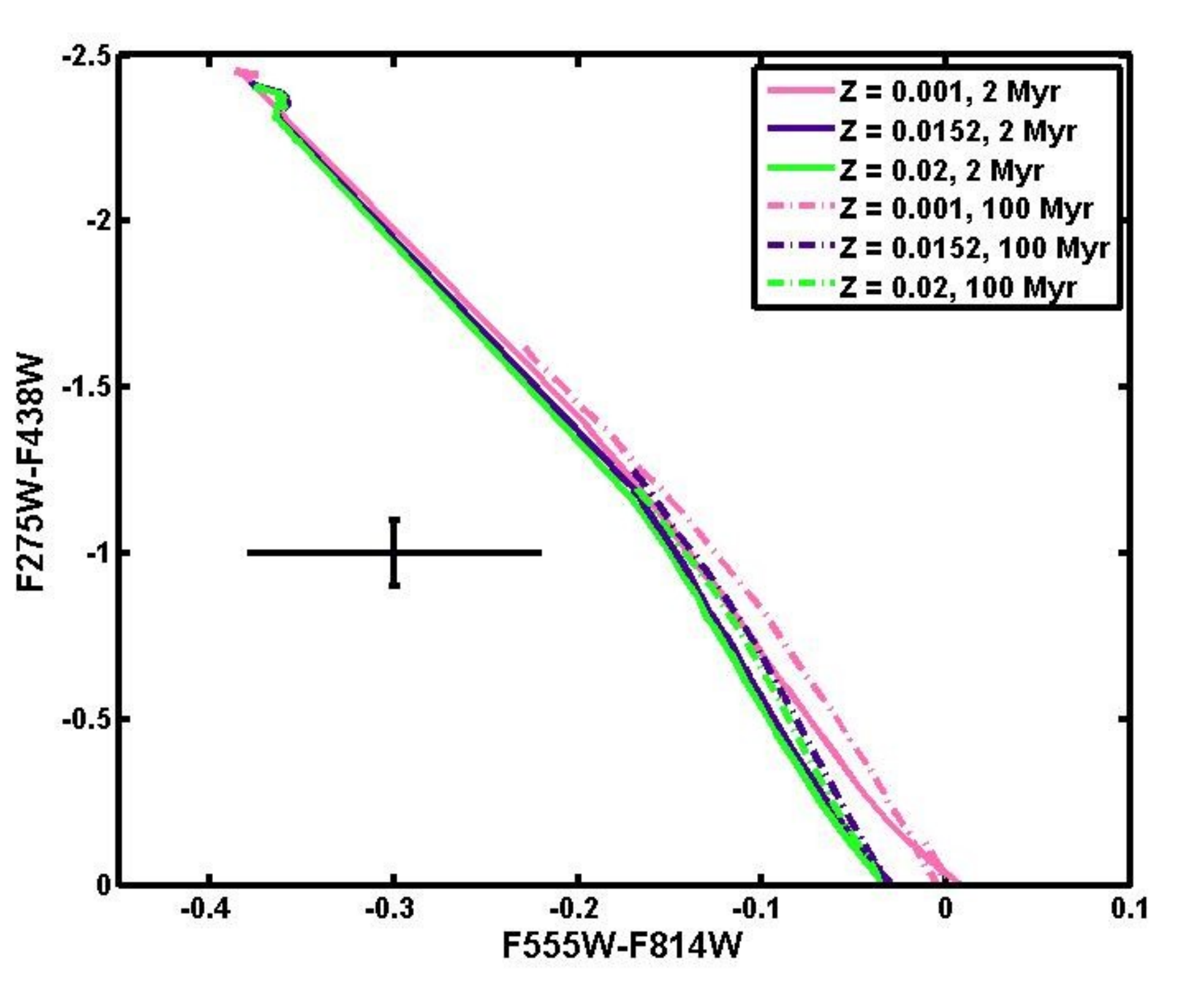}
            \caption{Color-color diagram of PARSEC isochrones of 2 and 100 Myr with varying metallicities 
   	  		\citep{parsec}. Solid lines are for 2 Myr, dashed-dot lines are for 100 Myr. Dark blue is for solar metallicity,
   	  		green is for higher metallicity, and pink is for lower metallicity. The black errorbars to the bottom left show the 
   	  		typical photometric error from LEGUS. Isochrones are for stars with masses greater than 2 M$_\odot$.}            
            \label{fig:metals}
          \end{figure}

          Unlike for the red clump method, there is no significant shift in the isochrones due to metallicity or age, 
          so our extinction correction should not be age- or metallicity-dependent, as seen in Figure 
          \ref{fig:metals}. For clarity, we only show the 2 Myr and 100 Myr isochrones, 
          as intermediate ages all lie in the same location. Note that the scale on Figure \ref{fig:metals} is zoomed 
          in to make the small shift in the isochrones visible, 
          thus the errorbar appears larger than in Figure \ref{fig:ccdcorr}.         

          \cite{reines10} find a significant contribution from emission lines for both the F555W and F814W HST filters for
          young massive clusters surrounded by HII regions. Since we focus on young massive stars, it is possible that 
          our stars are surrounded by or near HII regions, and therefore may suffer some of this contamination. However, 
          unless the HII region is an unresolved point source directly on top of an individual star, it is unlikely to be
          classified as a star by DOLPHOT if the contamination is significant. In the worst case, \cite{reines10} find that 
          about half of the flux in the F555W filter is from emission line contributions, while F814W suffers from $\sim 25 \%$
          emission line contribution. Combined, this results in a shift of the observed $(V-I)$ color of $\sim 0.2$, which will
          increase the observed Q by $\sim 0.3$. These are close to our photometric errors, and the overall effect will be to 
          slightly lower the derived $A_V$ for the affected stars. 
          
          It should also be noted that \cite{mason01} find a high incidence of binary star systems among young massive stars.
          Since this extinction correction relies on a star's position in color-color space, we should be largely unaffected.
          A binary system with one high-mass and one low-mass star will have the color of the high-mass star, since the low-mass
          star is too faint to significantly affect the SED of the whole system. The overall flux increases, but color (and therefore
          the Q-value) remains the same. Thus, we can generate accurate extinction corrections for stars in the LEGUS catalog even
          if they are blends of multiple stars. 
          
      \section{Completeness}
      \label{sec:complete}
          In this study, we only included stars more massive than 2 M$\odot$, since our extinction 
          correction method suffers from the isochrone degeneracy issue for stars less massive than
          this limit. 
		  Theoretically, using the NUV filter
          allows us to correct all stars with a $Q_{NBVI} \leq 0$, however practically,
          photometric uncertainties in the LEGUS sample require us to make an additonal
          apparent magnitude cut of $m_{\mathbf{NUV}} \leq 25$ and $m_{B} \leq 26$ before generating 
          an extinction correction. 
          
          There is an inherent completeness issue in our analysis, in that fainter, lower 
          mass stars cannot be observed at high extinction. Without applying a mass cut in addition to
          our apparent magnitude cutoff, we will tend to underestimate extinction, since we simply do not 
          see the lower mass stars that have high extinctions.  Table \ref{tab:masslimits} shows the minimum
          mass stars that can be corrected for a galaxy given our apparent magnitude limits, assuming no extinction. 
          Table \ref{tab:extinctlimits} shows the minimum mass of main sequence stars that can be detected 
          at $A_V = 0$, $A_V = 1$, $A_V = 2$, and $A_V = 3$ for each galaxy.
		  
		  \begin{deluxetable}{cccccc}
			\tablewidth{\textwidth}
			\tablecolumns{6}
			\tablecaption{Absolute magnitude, stellar mass, and $Q_{NBVI}$ limits for our 
			sample for apparent magnitude limits of 25 in F275W.\label{tab:masslimits}}
			\tablehead{
				\colhead{Galaxy}  
					& \colhead{Foreground $A_V$\tablenotemark{a}}
					& \colhead{M$_{275}$\tablenotemark{b}} 
					& \colhead{M$_\ast$\tablenotemark{c}} 
					& \colhead{$Q_{NBVI}$\tablenotemark{c}}\\
				\colhead{} & \colhead{} & \colhead{} 
					& \colhead{(M$_\odot$)} & \colhead{}
				}
			\startdata
			NGC 628	 & 0.196	   &	-5.37				&	14 & -1.85	 \\
			NGC 6503 & 0.084   &	-3.78				&	8  & -1.61	 \\
			NGC 7793 & 0.054   &	-2.79				&	6  & -1.43	 \\
			UGC 4305 &  0.175  &	-2.60				&	5  & -1.29	 \\
			UGC 5139 & 0.141   &	-3.28				&	7  & -1.51	 \\
			\enddata
			\tablenotetext{a}{Foreground extinctions from \cite{schlafly11}.}
			\tablenotetext{b}{Absolute magnitude of a star of apparent magnitude 25 in the 
				F275W filter, given a galaxy's foreground extinction and distance and 
				assuming local $A_V = 0$.}
			\tablenotetext{c}{Approximate stellar mass and $Q_{NBVI}$ of a star of
				apparent magnitude 25 in the F275W filter, given a galaxy's foreground
				extinction and distance and assuming local $A_V = 0$. Both values taken
				from a 2 Myr PARSEC ischrone \citep{parsec}}. 
		\end{deluxetable}
		
		Table \ref{tab:extinctlimits} also implies that for more nearby galaxies, we are seeing a slightly 
		different population of stars, as less massive stars become visible. Thus, to ensure we are measuring the 
		dust--to--gas ratio across a consistent stellar population, we apply an absolute magnitude cut of -5 in the 
		NUV-band. From the PARSEC isochrones, this corresponds to an approximate mass of 14 
		$M_\odot$ when combined with our $Q_{NBVI} \leq 0$ cut.
		
		\begin{deluxetable}{ccccc}
			\tablewidth{\textwidth}
			\tablecolumns{5}
			\tablecaption{Minimum stellar masses that can be corrected at given levels of extinction with an apparent
				magnitude limit of 25 in the F275W.\label{tab:extinctlimits}}
			\tablehead{
				\colhead{}		 & \multicolumn{4}{c}{Mass Limit ($M_\odot$)} \\
				\colhead{Galaxy} & \colhead{$A_V = 0$} & \colhead{$A_V = 1$} & \colhead{$A_V = 2$} & \colhead{$A_V = 3$} 
				}
			\startdata
			NGC 628		&	14			&	35			&	$>$100		&	 -			\\
			NGC 6503	&	8			&	16			&	45			&	$>$100		\\
			NGC 7793	&	6			&	11			&	24			&	70			\\
			UGC 5139 (Ho I)	&	7		&	14			&	35			&	100			\\
			UGC 4305 (Ho II)&	5		&	11			&	24			&	70			\\
			\enddata
			\tablecomments{We limit the PARSEC 
			\citep{parsec} isochrones to stars of less than 100 M$_\odot$, so we do not 
			include a limit for a given extinction if it is beyond that mass. Mass limits
  			in the table have been calculated including the effect of foreground
  			extinction to individual galaxies in the sample.}
		\end{deluxetable}
		
\begin{figure*}[h!]
          	\centering
          	\includegraphics[width=1.0\textwidth]{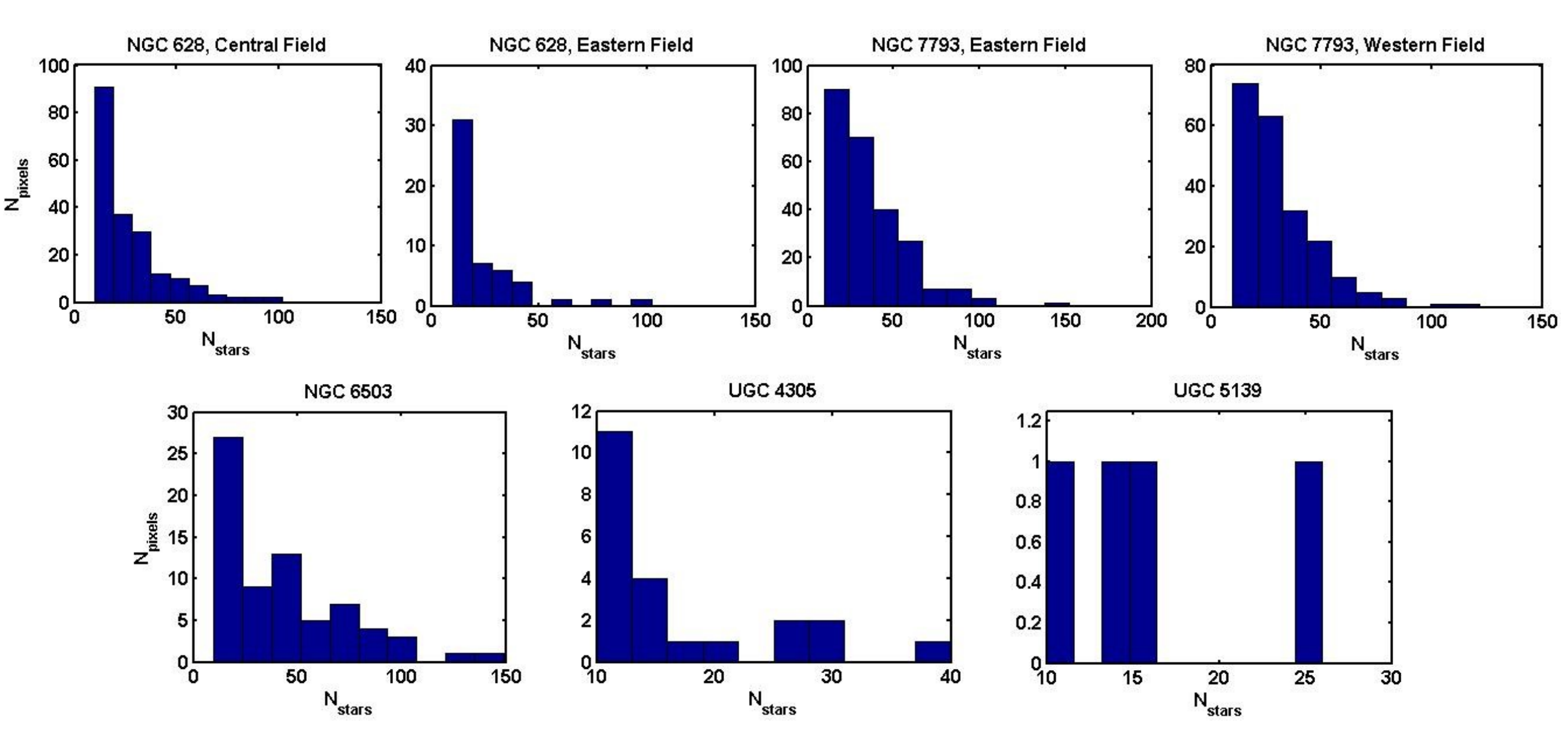}
          	\caption{Top row, left to right: NGC 628 central field,  NGC 628 eastern field,
          	NGC 7793 eastern field, NGC 7793 western field. Bottom row, left to right: NGC 6503, UGC 4305, UGC 5139.
          	Each plot shows a histogram of the number of pixels with a given number of stars in 10"x10" extinction 
          	maps in each pointing.} 
          	\label{fig:numberhists}
          \end{figure*}

		This higher mass cutoff, however, could create a geometric completeness problem, as we limit the number of
		usable stars significantly. In Section \ref{sec:gascorrs}, we require a minimum of 10 stars per pixel
		in our extinction maps to adequately sample the extinction in that pixel. The reasons for this are 
		twofold: first, a higher number of stars will more accurately sample the line-of-sight, as a higher number of stars is more likely
		to sample the entire dust layer. Second, a lower resolution better samples the resolution
		seen by the neutral gas maps, which smear out gas filaments that might be fully resolved in a higher resolution
		map. Stars within the gas filaments likely have a higher extinction than those outside it. Thus, averaging over
		a similar resolution to the neutral gas maps helps minimize some of this potential variation, as it is more likely
		to find stars both in and out of gas filaments in the same area with a larger resolution. For determination of a 
		dust--to--gas ratio we must sample the stellar extinction over areas comparable to the resolution of the radio data. 
		After applying a magnitude 
		cut corresponding to approximately 14 M$_\odot$, we find that our analysis supports the construction of 10"x10"
		extinction maps for comparison to the neutral gas maps and calculate the dust--to--gas ratio using 10"x10" extinction maps. Figure
		\ref{fig:numberhists} shows histograms of the number of pixels with the required number of stars in each galaxy.
		
		This introduces a potential bias towards lower extinctions, in that we exclude regions
		of high extinction where we cannot observe many stars because they are obscured. We test 
		the effects of including pixels with fewer stars in Section \ref{sec:dgrvmetal}.
		
		It should also be noted that the version of the LEGUS catalogs used in this paper is conservative 
		and includes only stars with the most accurate photometry. This restriction, discussed in \cite{sabbi17}
		excludes $\sim$10\% of stars in each pointing, so for our requirements, we miss perhaps $\sim$34\% of stars.
		This effect does not change the distribution of stars with respect to apparent magnitude, only the total
		number of stars, so we do not expect a significant change in our results. The conservative catalog
		is sufficient to support the development of the extinction maps in Section \ref{sec:extinctmaps}. 		 
      
      \section{Extinction Maps}
          \label{sec:extinctmaps}
          We generate extinction maps for each galaxy using the individual stellar extinctions. Stars are
          binned spatially, then a weighted average based on the uncertainty in an 
          individual star's extinction derivation is calculated for a given pixel. The quality of the
          map depends on both the resolution and the number of stars per pixel. In principle, we want high
          resolution, however, due to the uneven distribution of stars in galaxies, this is
          not always feasible. We also want a high number of stars in each pixel to reduce the
          effect of stars with high uncertainties on the derived extinctions and potential differences in line-of-
          sight distances. 
          
\begin{figure*}[h!]
          	\centering
          	\includegraphics[width=1.0\textwidth]{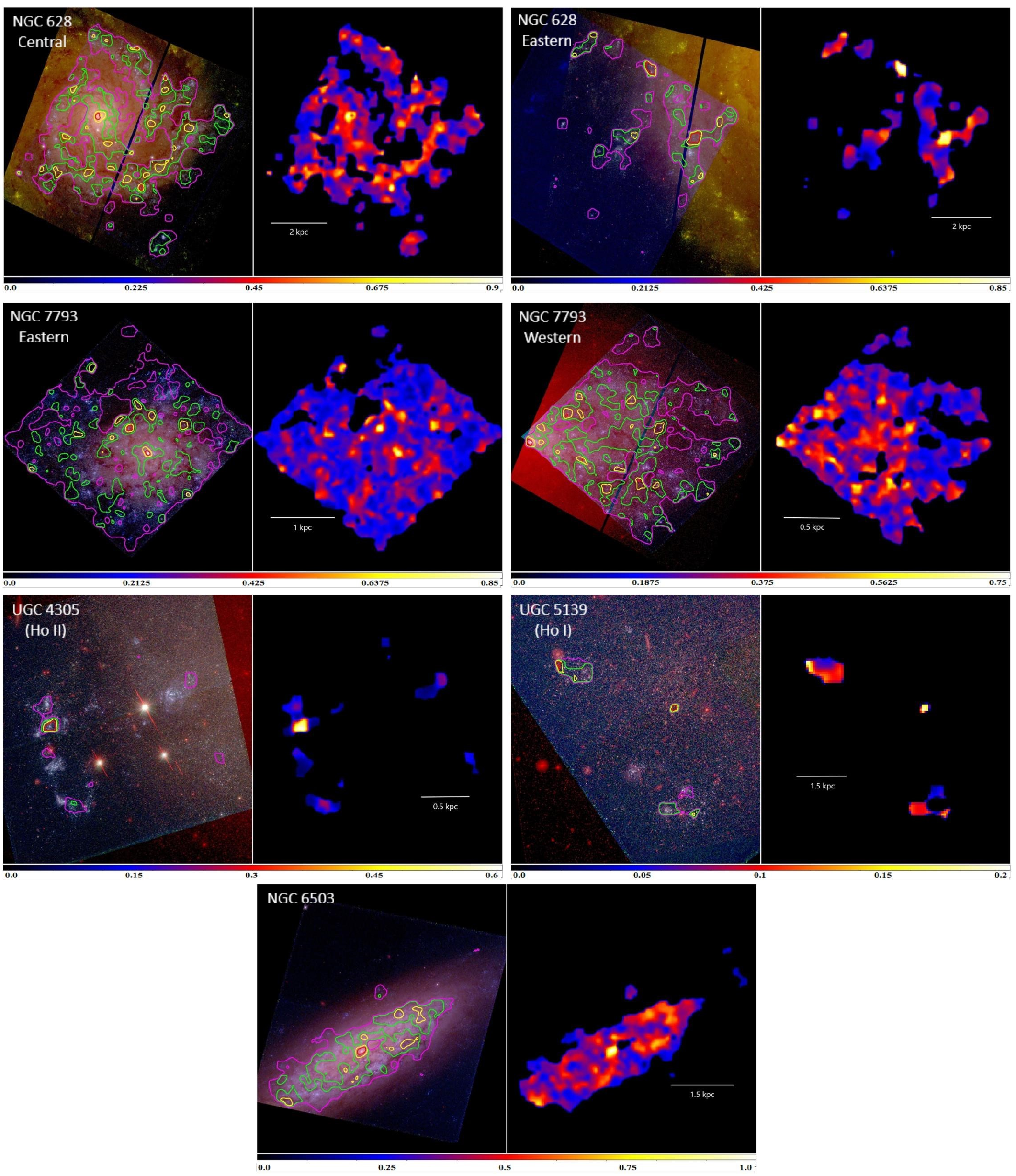}
          	\caption{Top row, left to right: NGC 628 central field,  NGC 628 eastern field.
          	Second row, left to right: NGC 7793 eastern field, NGC 7793 western field. Third row,
          	left to right: UGC 4305, UGC 5139. Bottom center: NGC 6503. The left image in each set is the 
          	3-color HST image, using I, B, and U-band images. Colored contours for the reddening are shown on 
          	the map to the right, with contour values equal to 20\% (magenta), 
          	40\% (green), 60\% (yellow), and 80\% (red) of the maximum extinction. 
          	The right image in each set is the weighted average reddening adaptive resolution map for each galaxy, 
          	smoothed with a 3" Gaussian. Maps show E(V-I) for each galaxy. Blue is low E(V-I), while yellow/white 
          	is high E(V-I) (see color bar). For a MW extinction law with $R_V = 3.1$, $A_V = 2.41E(V-I)$. }
          	\label{fig:alladapmaps}
          \end{figure*}

          To maximize resolution while maintaining enough stars for an accurate measurement of the 
          extinction, we employ an adaptive resolution algorithm. In Figure \ref{fig:alladapmaps}, we show 1" sampled 
          adaptive resolution extinction maps for each of 
          our fields. Starting in the upper left, we define the smallest possible region that contains 10 stars. 
          The smallest region we consider is 1"x1". If a region does not contain 10 stars, we move on to a 2"x2", 
          5"x5", and 10"x10" region, stopping when we have at least 10 stars and using the starting 1"x1" area as
          the center. If there are not 10 stars available
          even in the 10"x10" case, we do not calculate an extinction for that pixel. 
          We then repeat this for the 
          next pixel position, moving across the row. At the end of the row, we move 1" down and 
          repeat the process. For example, in the most crowded regions, regions with an adequate
          number of stars have a pixel size of 1"x1", however on the outer edge of the galaxy, we may be required to 
          sample stars upwards of 5" from the central point of a pixel in order to meet the minimum required number 
          of stars. The effective resolution of these regions is then 10"x10". This sometimes results in counting the same
          star in multiple pixels in the sparsest regions, though the effect is to spread the extinction out over a larger
          area as opposed to skewing the extinction to a higher value. This adaptive map is smoothed using a 3" Gaussian to 
          remove the sharper discontinuities in some regions.
          
          In general, the dwarfs UGC 5139 and UGC 4305 have much lower extinctions than NGC 628 or NGC 
          6503. We also see good correlation between areas of high extinction in the extinction maps and visible
          patches of dust in the optical image, such as the center of NGC 628, and the edges of visible dust lanes in the 
          spiral arms of NGC 628 and the edge of NGC 6503. However, we also see large
          empty regions in the middle of the maps, particularly for NGC 628. These are 
          regions where we have a lack of coverage of stars, not necessarily regions
          with no extinction. Some, like the very large dust lane in NGC 628, may be 
          regions of extremely high extinction, where stars are dimmed below our 
          ability to generate accurate photometry. Young massive stars tend to be observed offset from these regions, as
		  density-wave progression implies that these stars appear most often downstream from the main dust lanes in 
		  spiral arms \citep{wielen74}. \cite{elmegreen14} did find that there are massive stars in dust lanes, however 
		  they cannot be observed in the optical/NUV, as they are highly obscured. We also have the problem of crowding. 
		  For the most crowded regions of the galaxies, point sources are blended together and therefore not resolved. 
		  These sources then cannot be extracted by DOLPHOT and are not in our catalogs.
          
          Table \ref{tab:extinctlimits} implies that for more distant galaxies, we are unable to measure extinctions
		of more than 2 magnitudes in the V-band. However, as seen in Figure \ref{fig:alladapmaps}, we do see
		extinction map pixels for NGC 628 in the central field with $A_V > 2$. These regions are near the crowded
		center of the galaxy, so it is possible that we are seeing a high percentage of blends in this particular
		case. Blends in more distant galaxies may still appear as point sources and are extracted by DOLPHOT as 
		individual stars when there may be two, three, or even more stars contributing to the light seen from a 
		source. Some may even be very low-mass clusters, leading to a higher--than--expected magnitude for a given
		color. The extinctions in this crowded region are likely to be unreliable, and should therefore
		be treated with the appropriate caution.
		
		In principle, these maps could be used for correcting H$\alpha$ and optical images for extinction. 
		They also are useful for qualitative comparisons to emission maps of dust from IR and FIR imaging. In both
		cases, we recommend caution, as these maps are likely lower limits for the extinction in these galaxies. We 
		show them here for illustrative purposes, however the analysis in sections \ref{sec:gascorrs} and 
		\ref{sec:discussion} relies on 10"x10" maps.
		          
    \section{Dust--to--Gas Ratios}
        \label{sec:gascorrs}          
        The total neutral hydrogen gas column density depends on both the HI and H$_2$ column density. The HI 
        column density can be taken directly from 21 cm flux maps (assuming optically thin HI), but
        the H$_2$ column density must be calculated from CO flux maps using an assumed
        $X_{CO}$. For these maps, we use the same X$_{CO}$ used by \cite{rahman11} and \cite{leroy09} of
        $2 \times 10^{20}$ cm$^{-2}$ (K km s$^{-1}$)$^{-1}$. The total neutral hydrogen column density in
        a given line of sight is then
        
        \begin{equation}
            N_H = N_{HI} + 2N_{H_2}.
            \label{eq:NH}
        \end{equation}
        
        From \cite{rachford09} and \cite{bohlin78}, for sightlines with a standard Milky Way 
        extinction law, one finds equation \ref{eq:dmratioext}. We can test if this relation applies 
        to the galaxies in this paper by spatially correlating the neutral gas maps with the dust maps 
        generated in Section \ref{sec:extinctmaps}. The adaptive resolution maps are useful for qualitative 
        comparisons to the location of spiral arms and dust lanes in
          the galaxy, however the radio data are also at a
          much lower resolution than the adaptive resolution maps. The neutral gas is
          therefore averaged out over a larger area, which leads to geometric effects
          if the resolutions are not properly matched. To compare the extinction maps to 
          the neutral gas maps, we simply bin the stars spatially
          into 10''x10'' bins and calculate weighted averages of the extinction for each bin.
          
\begin{figure}[h!]
          	\centering
            \includegraphics[width=0.5\textwidth]{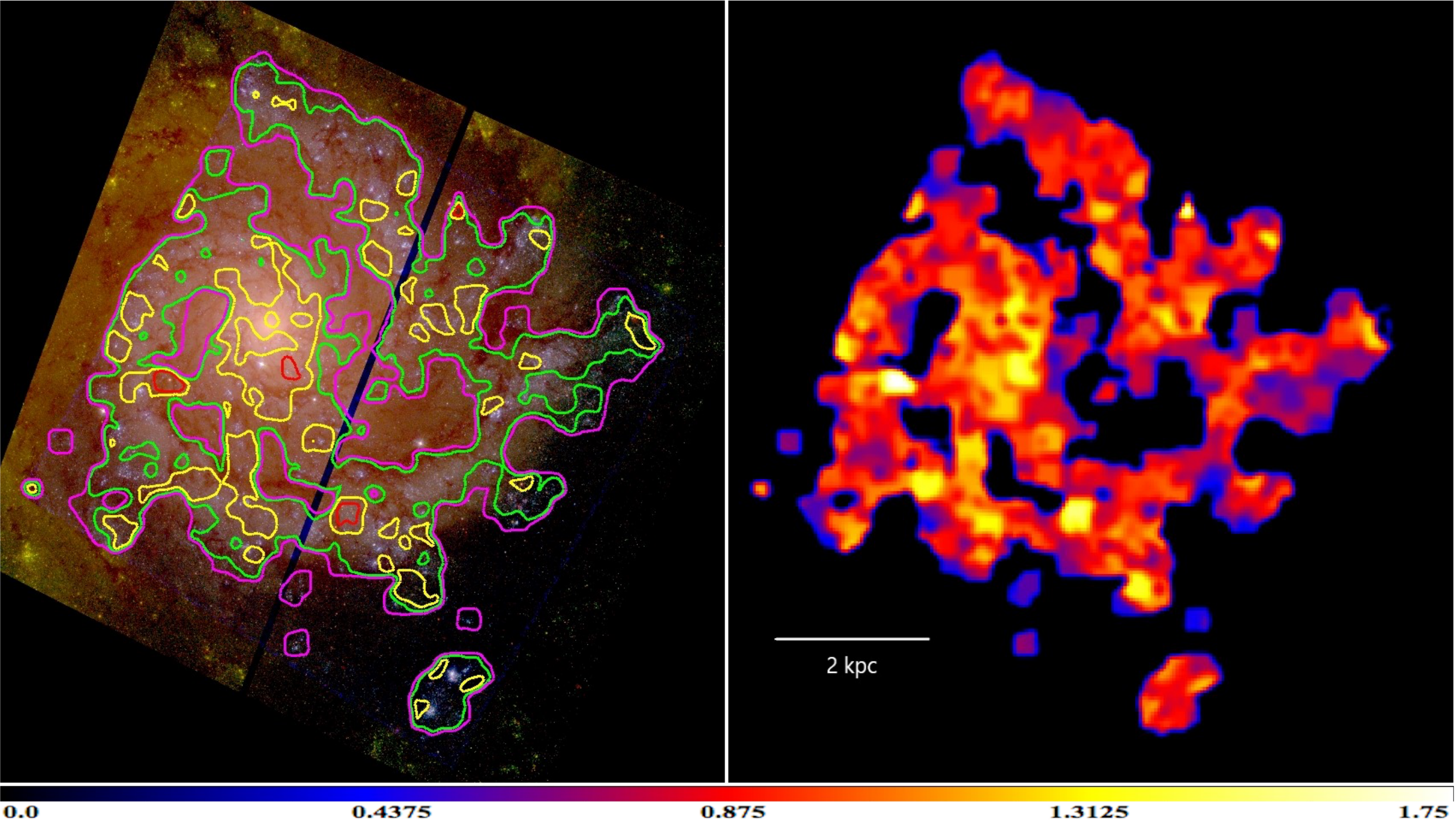}
            \caption{Central field of NGC 628. Left: 3-color HST image, using F814W, F435W, and F336W. Colored 		
            contours are at 20\% (magenta), 40\% (green), 60\% (yellow), and 80\% (red) of the maximum extinction 
            shown in the map to the right. Right: Adaptive resolution maximum extinction map of 
            the central field of NGC 628. Blue is low E(V-I), while yellow/white is high E(V-I) (see color bar). 
            For a MW extinction law with $R_V = 3.1$, $A_V = 2.41E(V-I)$.}
            \label{fig:adap628cmax}
          \end{figure}

          We also need to consider that since we are using exclusively young stars, which 
          usually lie in the midplane of the disk, this
          only represents half the gas column density along a line of sight. Therefore, we 
          may need an estimate of the maximum extinction derived for individual stars along the line--of--sight. 
          Since the most highly obscured stars in each bin are also likely the faintest stars, and therefore stars with
          the highest photometric uncertainties, a standard maximum is not necessarily representative of the
          maximum value of the pixel. Thus, we calculate the weighted average of the upper quartile of the
          stars in the bin as the maximum of that bin. This ``maximum" is shown in Figure \ref{fig:adap628cmax} 
          for the central field of NGC 628. We also require that all pixels in the extinction map 
          contain at least 10 stars, to ensure that we are properly sampling the line of sight of a pixel. 
          The upper quartile therefore always consists of at least the 3 most-obscured stars. Both the
          weighted average and the ``maximum" values for extinction are used to quantify the effect of
          estimating $A_V$ in this way in the analysis that follows.
          
           %MAXIMUM EXTINCT MAPS
          
\begin{figure*}[h!]
        	\centering
            \includegraphics[width=\textwidth]{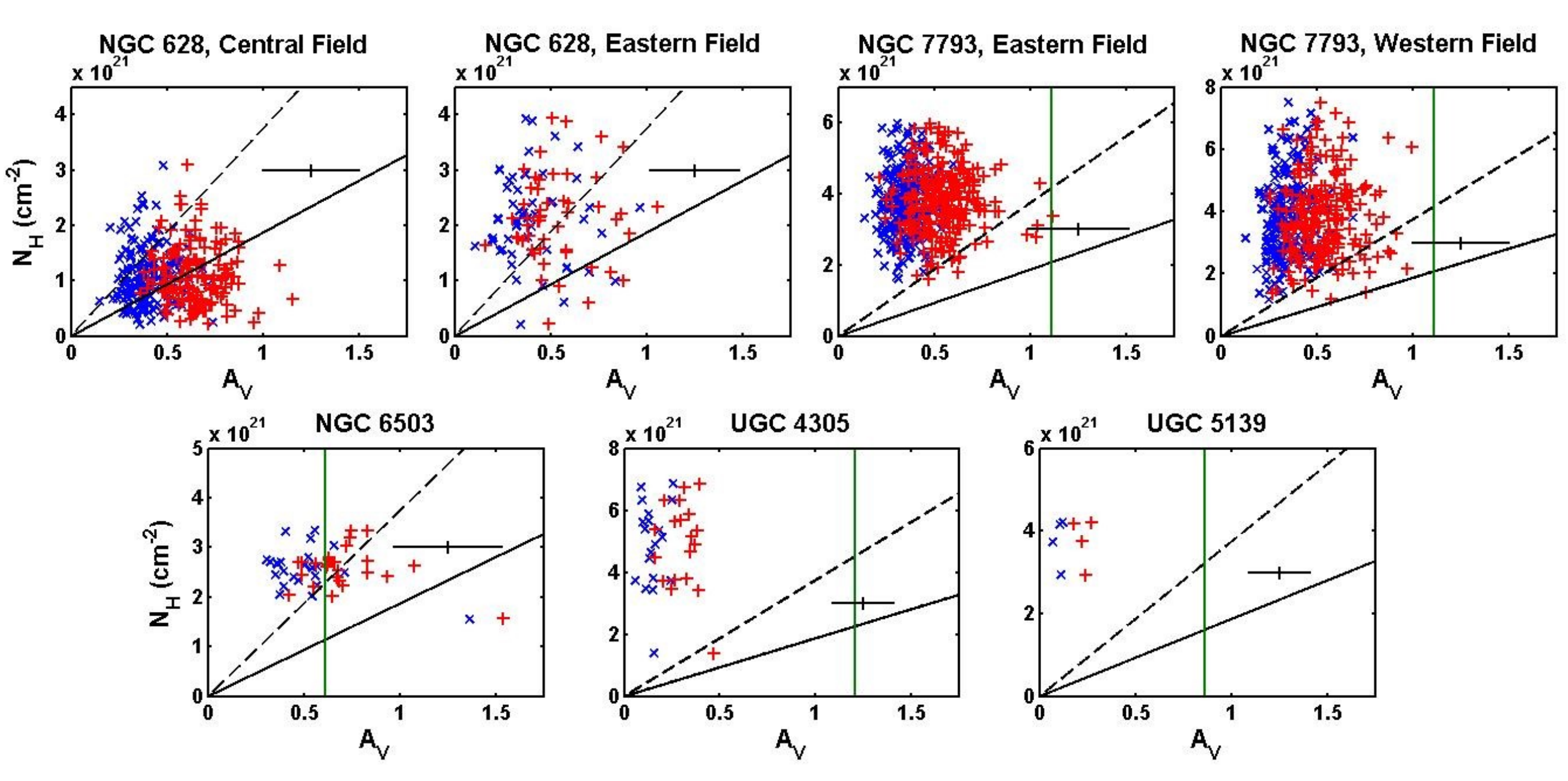}
            \caption{Extinction vs. total neutral gas column density for all galaxies. Blue points 
            are extinctions from the weighted average 10" extinction maps, red are from the 
            ``maximum" 10" maps, where the average of the upper quartile of the extinction is used 
            for each pixel. Dashed lines are for the half--MW relation from equation
            \ref{eq:halfgasdustcorr}, while the solid lines are the full--MW 
            relation from equation \ref{eq:dmratioext}. The black horizontal errorbar shows the typical rms of individual 
            stellar derived extinctions within a pixel. The green vertical line shows
            the maximum observable extinction for a star of absolute magnitude -5 in
            the F275W filter, after correcting for foreground extinction. The two plots
            for NGC 628 do not show a green line because the faintest star that can be 
            observed with $A_V = 0$ has an F275W magnitude of -5.37.
            Top row (left to right): Central field of NGC 628, eastern field of NGC 628, 
            eastern field of NGC 7793, western field of NGC 7793. Bottom row (left to 
            right): NGC 6503, UGC 5139 and UGC 4305.}
            \label{fig:gascorrs}
        \end{figure*}

          The recent PHAT survey used the assumption of a log-normal distribution of reddenings along
          the line of sight to generate 
          their extinction maps \citep{dalcanton15} from individual red giant stars. This more accurately models a thin dust screen. The 
          population of stars that the PHAT survey is sensitive to is much older than 
          the LEGUS catalog. Thus, the potential for stars to exist significantly above
          or below the midplane of the disk (and therefore the majority of the dust)
          is much higher. \cite{dalcanton15} estimate that the scale height of M31's stars to be more
          than a factor of 10 times the likely depth of the dust, thus a thin screen with a log-normal 
          distribution of dust is more appropriate for their sample. The vast majority of young massive stars
          are likely distributed with a scale height similar to that of the dust of the disk (and
          within the dust layer). Since our sample is limited to young massive stars, we may assume a simpler
          model of a smooth layer of dust within a stellar disk. We expect a symmetric 
          histogram of extinction values around a mean for each 10" pixel, as long as higher extinctions are not so
  		  high that we run out of highly obscured stars. On
          average, we may expect a typical young star to be in the middle of the disk plane and 
          therefore probe half the extinction implied by the total gas column, in which case
          the expected relation would be: 
        
        \begin{equation}
            \frac{A_V}{N_H} = 2.65 \times 10^{-22} \:\mathrm{mag \:cm^2\:H^{-1}}.
            \label{eq:halfgasdustcorr}
        \end{equation}
        We show the full--plane and half--plane relations for the MW from equations \ref{eq:dmratioext} and 
        \ref{eq:halfgasdustcorr} in Figure \ref{fig:gascorrs}.
        
        The HI and CO data are mostly taken from the THINGS, LITTLE THINGS, and HERACLES
        surveys, and supplemented where necessary by other sources. The neutral gas map
        data sources and beam sizes are summarized in Table \ref{datasources}.
        
        \begin{deluxetable*}{ccccc}
        	\tablewidth{0.5\textwidth}
        	\tablecolumns{5}
        	\tablecaption{Summary of Neutral Gas Map Data \label{datasources}}
        	\tablehead{
        		\colhead{Galaxy} & \colhead{HI Source\tablenotemark{a}} 
        		& \colhead{HI Beam Size} & \colhead{CO Source\tablenotemark{b}}
        		& \colhead{CO Beam Size}\\
        		}
        	\startdata
        	NGC 628 		& THINGS 			& 11.88"x9.3" 	& HERACLES 		& 11" \\
        	NGC 6503 		& \cite{greisen09} 	& 14" 			& CARMA-STING 	& 4.3"\\
        	NGC 7793 		& THINGS 			& 15.6"x10.85" 	& ... 			& ...\\ 
        	UGC 4305 (Ho II)& LITTLE THINGS 	& 13.74"x12.57" & HERACLES  	& 11"\\
        	UGC 5139 (Ho I) & LITTLE THINGS 	& 14.66"x12.73"	& HERACLES  	& 11"\\
        	\enddata
        	\tablenotetext{a}{HI data sources. The HI Nearby Galaxy Sample (THINGS) is described in \cite{walter08}.
        		Local Irregulars That Trace Luminosity Extremes (LITTLE) THINGS is described in \cite{hunter12}.}
        	\tablenotetext{b}{CO data sources. The HERA CO-Line Extragalactic Survey (HERACLES) is described in 
        	\cite{leroy09}. The CARMA Survey Toward IR-bright Nearby Galaxies (CARMA-STING) is described in 
        	\cite{rahman11}.}
        \end{deluxetable*}
        	       	        	
        Figure \ref{fig:gascorrs} shows the relation between inferred extinction and
        gas column for 10"x10" pixels. There is significant scatter in each plot, 
        suggesting influence from geometric effects caused by variations of the 
        location along the line-of-sight of stars in a pixel. On all of the plots, the vast
        majority of points lie to the left of the standard Galactic dust--to--gas ratio, 
        as expected. We are unlikely to observe the full column density of dust, as we are 
        unlikely to observe stars on the far side of the disk. In fact, the high number of 
        points to the right of both the half and full MW relations in NGC 628's central field may imply a need
        for a higher dust--to--gas ratio in that region. NGC 628 and NGC 6503 both have most of their scatter
        around either the half-- or full--MW relation. NGC 7793 and two dwarfs, on the other hand, 
        show almost vertical relations, with little to no points at high extinctions for
        high dust column densities, implying a smaller dust--to--gas ratio. Given our limits shown in 
       	Table \ref{tab:masslimits}, we certainly could have observed higher extinctions for the higher
       	gas column densities, especially for the dwarfs, but we do not.
        
\begin{figure}[h]
        	\centering
            \includegraphics[width=0.5\textwidth]{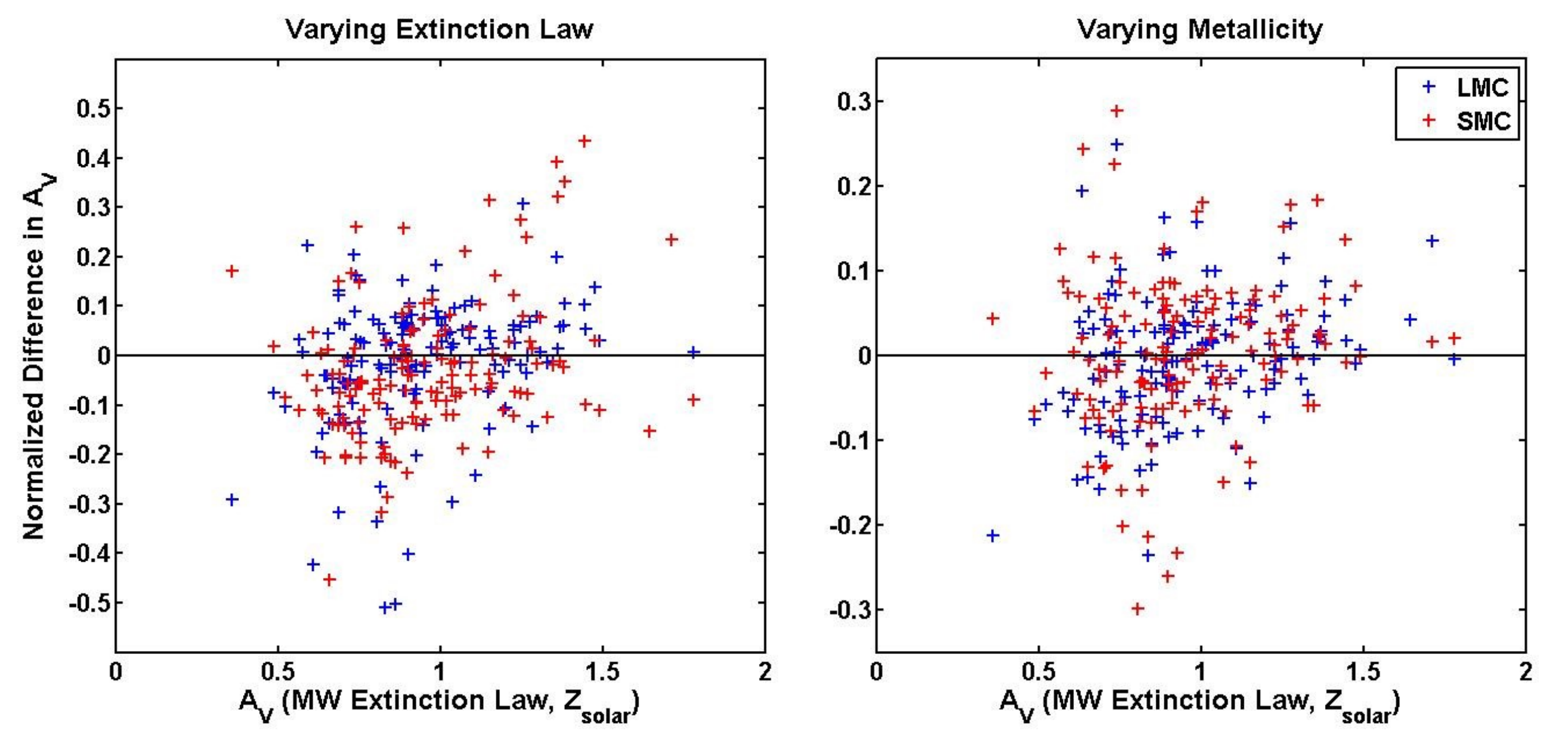}
            \caption{$\frac{A_V(MW) - A_V(LMC)}{A_V(MW)}$  and $\frac{A_V(MW) - A_V(SMC)}{A_V(MW)}$
            for pixels in extinction maps of NGC 628's central field generated 
            by varying the extinction law used to determine the slope of constant 
            $Q_{NBVI}$ (left) and the metallicity of the isochrones (right). The black line in each plot marks 
            the zero line.}
            \label{fig:gascorrsdiffboth}
        \end{figure}

        The assumption of solar metallicity and a Milky Way extinction law may not hold for 
        all galaxies. The correction in Section \ref{sec:extinctcorrs} assumes both a MW reddening law and a solar 
        metallicity set of isochrones. Individual stars may show a large variation, but these are averaged out, so 
        the overall effect on 
    the final maps is small. Varying the extinction law to that of the LMC or the SMC \citep{gordon03} causes an average normalized difference of about 1-2\% in 
    the weighted average extinction across the 
    map, with an RMS variation of about 13\%. The effect from varying the metallicity to the metallicity
    of the young population of the LMC or SMC is smaller, at about 0.5\%, with
    an RMS variation of 7-9\%. Figure \ref{fig:gascorrsdiffboth} shows the effect of varying extinction law and metallicity
    on the derived $A_V$ for extinction maps in NGC 628. Most pixels remain within about 15\% for a change in extinction law, 
    and 10\% for a change in isochrone metallicity. Thus, the dominant effect on Figure 
        \ref{fig:gascorrs} appears to be from changes in the dust--to--gas ratio, not 
        changes in the local extinction law or metallicity.   

	\subsection{Radial Variations}
	\label{sec:radial}
\begin{figure}[h]
        	\centering
            \includegraphics[width=0.5\textwidth]{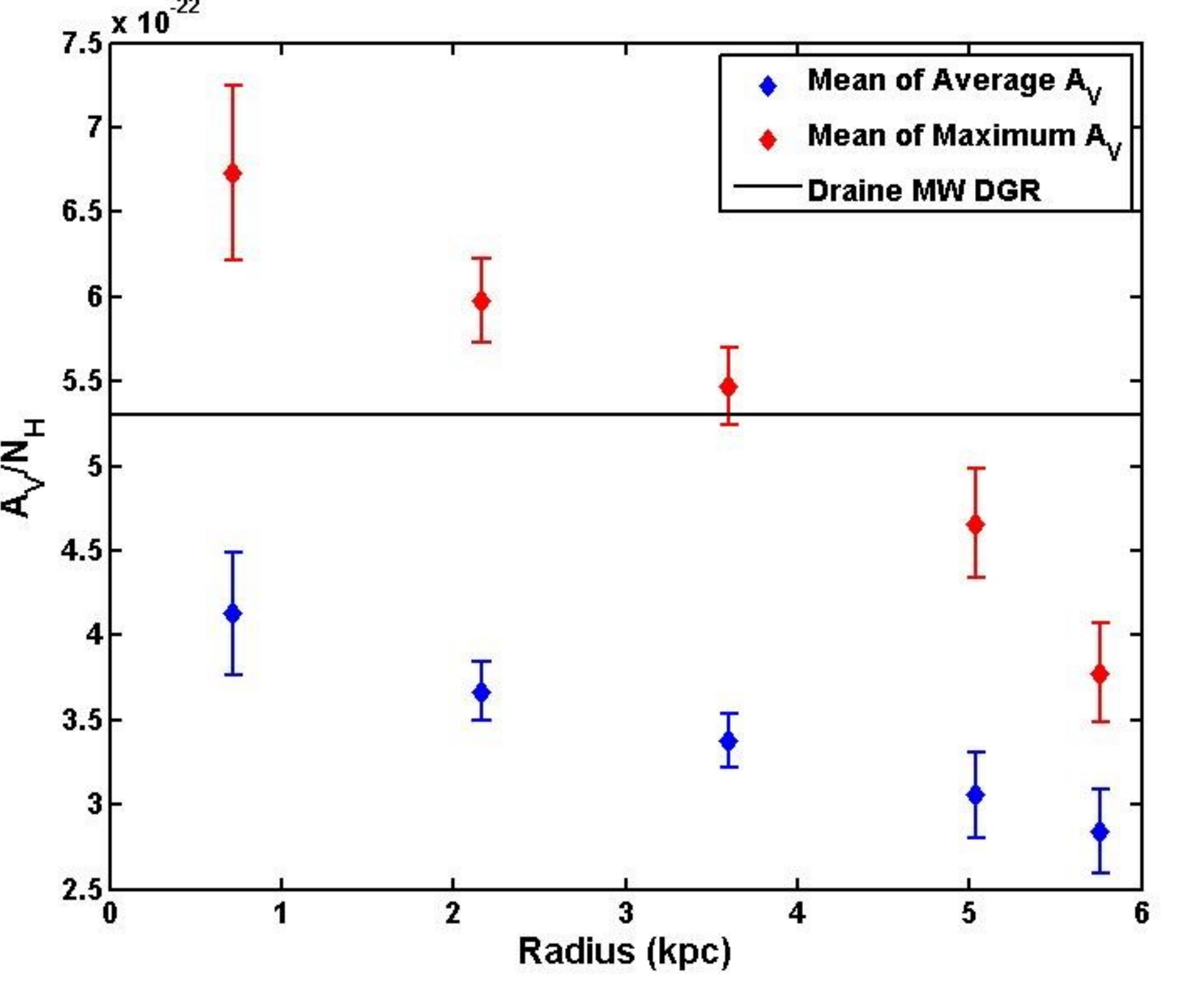}
            \caption{Average $\frac{A_V}{N_H}$ vs. galactocentric radius for NGC 628. Red diamonds use the ``maximum" extinction
            maps, while blue diamonds use the average extinction maps. The black line close to the top of the panel is the typically observed Milky 
            Way value. Errorbars are determined using the scatter of both $A_V$ and $N_H$ in that bin. Radial 
            distance is determined using the center of each bin, except for the outermost bin, which uses the 
            inner boundary.}
            \label{fig:avgdgrrad628}
    \end{figure}

	We combined the two fields of NGC 628, performed a radial deprojection, and binned pixel areas from the 
	extinction maps into five radial bins. The first four bins are 30" in size, which corresponds to a 
	physical size of 1.4 kpc. The outer bin encompasses everything not contained within the first four bins, and is,
	for our two pointings, similar in size to the other bins. 
	We took averages of the neutral gas column density and extinction 
	and calculated average dust--to--gas ratios in each bin. As seen in Figure \ref{fig:avgdgrrad628}, there is a clear
	decrease of dust--to--gas ratio with increasing radius in the outer bins, with a turnover in the inner bins.
	A fit to the ratios gives a power-law with a slope of $-0.031 \pm 0.004$, slightly shallower than the 
	observed metalliticy gradient of $-0.044$. All 
	of our values lie beneath the typical observed MW
	relationship, which is expected since we are likely not seeing the full column density of dust using our 
	method. 
	
    	It should be noted that NGC 628 is the farthest galaxy in our sample. At 9.9 Mpc, 
    	the HST pixel scale is 1.92 pc,
        which is likely insufficient to resolve individual stars, especially
        in crowded regions such as the galactic center. Since we still find extinction values that are consistent with 
        what might be expected from a MW relation despite this, we are likely seeing many blends in this galaxy in 
        particular. Blends can allow us to see high extinctions for pairs or groups of massive stars, as it 
        affects magnitudes, but blends with $Q_{NBVI}$ values in the range of massive stars do still seem to 
        enable us to derive consistent results in comparison to the other galaxies. Blends that produce redder Q's
        (like those produced by red supergiant stars) are excluded by our selection.
        
\begin{figure}[h]
        	\centering
            \includegraphics[width=0.5\textwidth]{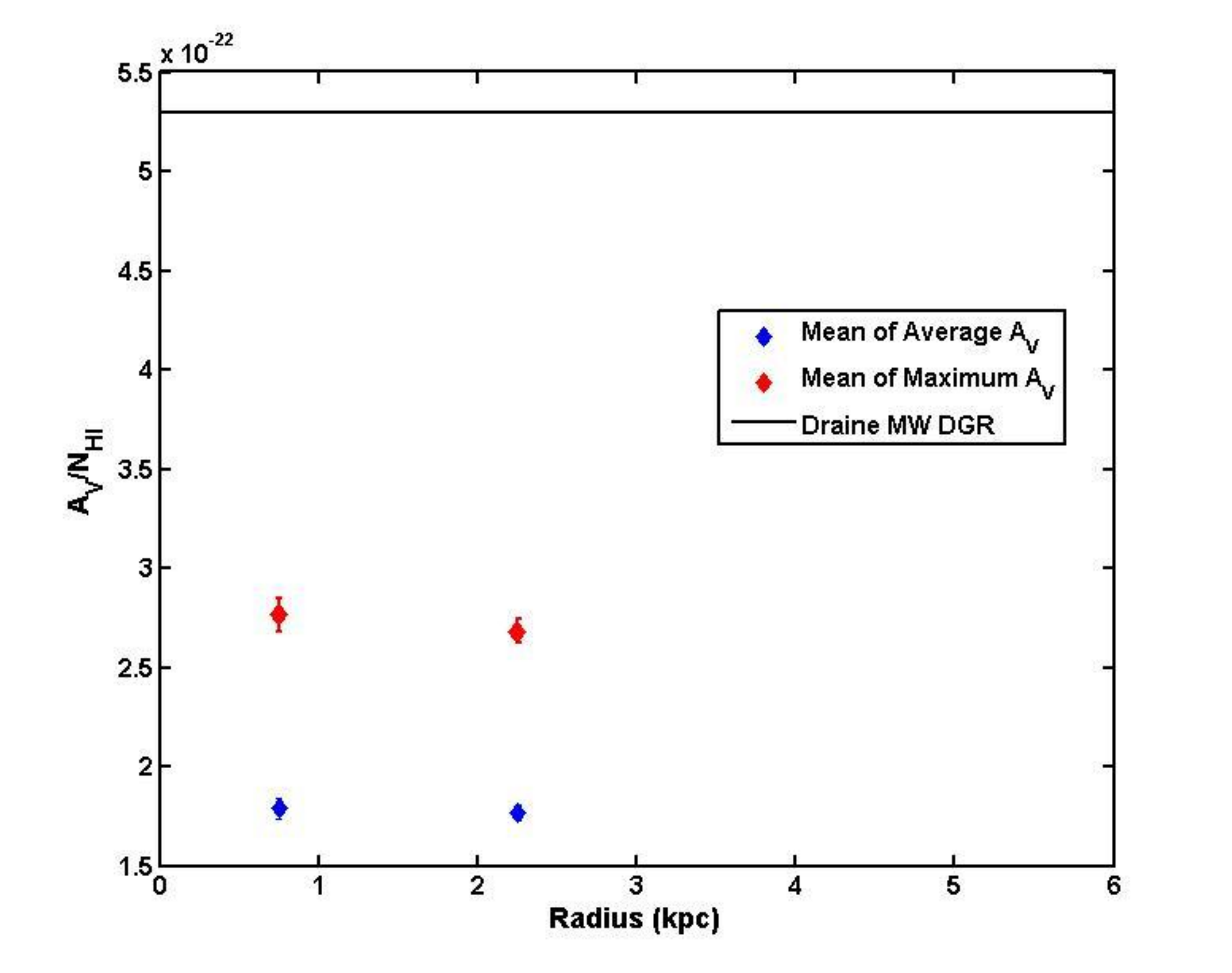}
            \caption{Average $\frac{A_V}{N_H}$ vs. galactocentric radius for NGC 7793. Red diamonds use the ``maximum" extinction
            maps, while blue diamonds use the average extinction maps. The black line is the typically observed Milky 
            Way value. Errorbars are determined using the scatter of both $A_V$ and $N_H$ in that bin. Radial 
            distance is determined using the center of each bin, except for the outermost bin, which uses the 
            inner boundary. Only pixels with 10 stars are included in this average.}
            \label{fig:avgdgrrad7793}
    \end{figure}

        We performed a similar radial analysis with NGC 7793 with bins of the same physical size to those in NGC 628. 
        NGC 7793 is a much smaller galaxy than NGC 628 in terms of physical size, so we were restricted to two bins instead 
        of five. As we show in Figure \ref{fig:avgdgrrad7793}, there is very little change in the dust--to--gas
        ratios between the two bins, as we might expect since we cover less than half the radial distance we do in NGC 628. 
        Both bins also have significantly lower dust--to--gas ratios than NGC 628, which we might expect from NGC 7793's 
        lower metallicity. As seen in Figure \ref{fig:gascorrs} and Table \ref{tab:extinctlimits}, it is possible to observe
        higher extinctions in NGC 7793, if they exist.
        
	\subsection{Variations with Metallicity}
	\label{sec:metallicity}
\begin{figure}[h]
        	\centering
        	\epsscale{2.3}
            \plottwo{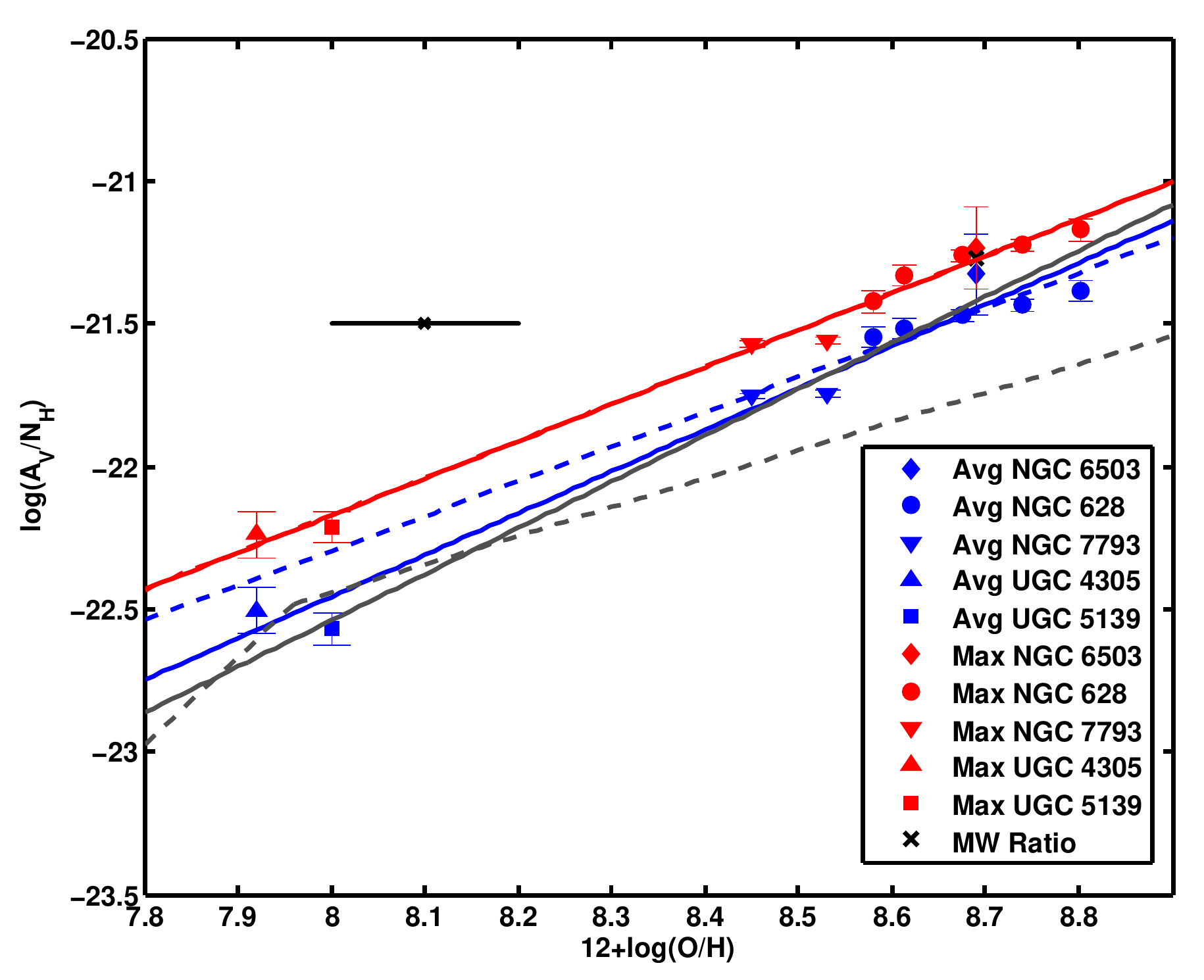}{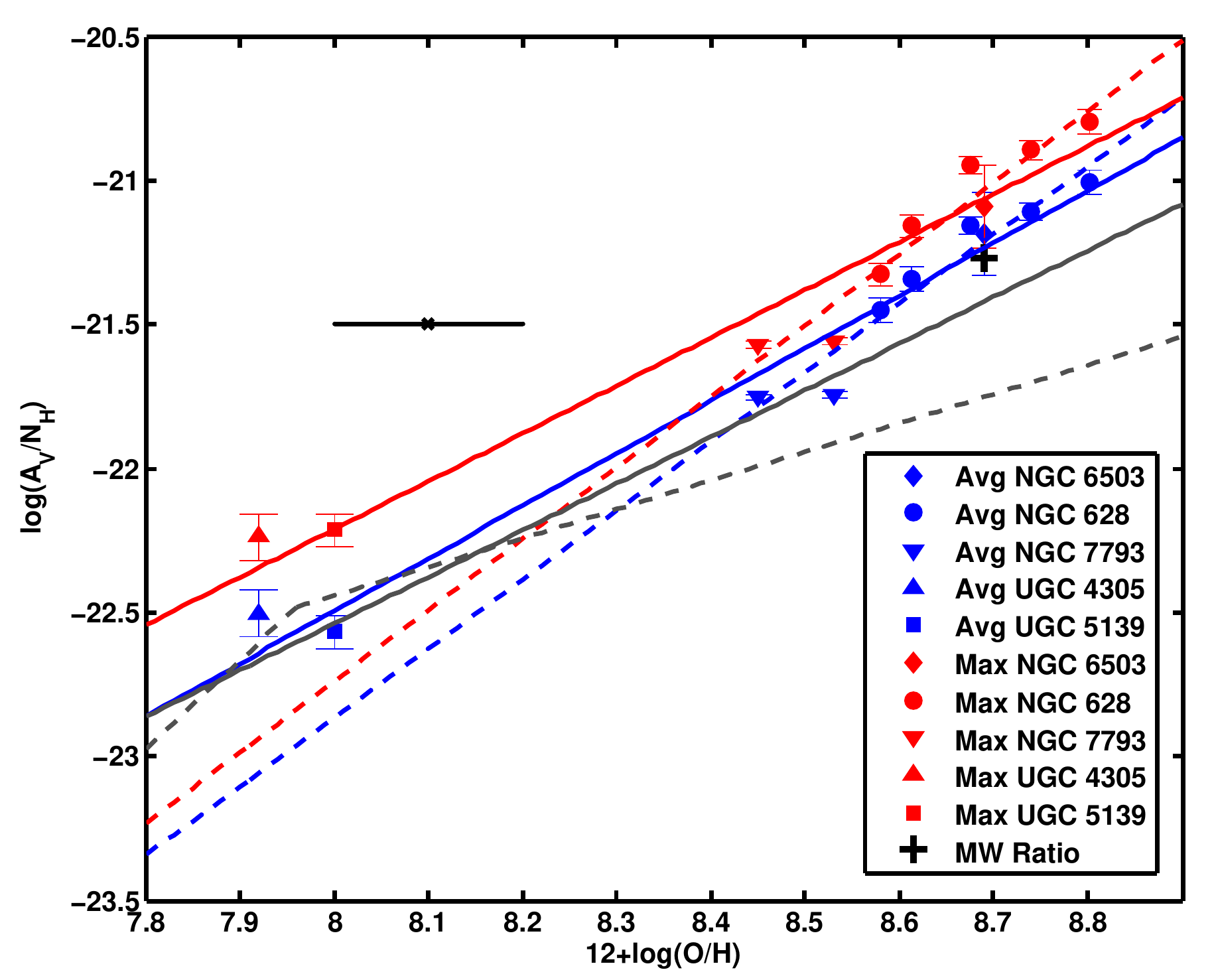}
            \caption{Average dust--to--gas ratio vs. metallicity for all NGC 628 radial bins, NGC 6503, NGC 7793 
            (both radial bins), UGC 5139, and UGC 4305. Red points use the ``maximum" extinction
            maps, while the blue ones use the average extinction maps. The black cross is the typically observed Milky 
            Way value. Vertical errorbars are determined using the scatter of both $A_V$ and gas column density 
            in that bin or galaxy. The black horizontal errorbar in the upper left shows the typical metallicity uncertainty
            from the literature. Dashed red and blue lines are fit to NGC 6503 and the NGC 628 radial bins, while solid red and blue lines
            lines are fit to the entire sample. The dark grey lines are the \cite{remy14}
            single (solid) and broken (dashed) power law relationships. Top: Dust--to--gas ratios calculated using $N_H = N_{HI} + 2N{H_2}$
            for all but NGC 7793. Bottom: Dust--to--gas ratios calculated with $N_{HI}$ only.}
            \label{fig:dgrmetal}
    \end{figure}

	Using the metallicity gradient for NGC 628 from \cite{berg12}, we assign a metallicity for each radial
	bin defined in Section \ref{sec:radial}. We do the same for the two bins in NGC 7793 using the metallicity gradient
	from \cite{stanghellini15}. For NGC 6503, UGC 4305, and UGC 5139, we calculate the dust--to--gas ratio for the whole galaxy, since 
	there are either no published metallicity gradients (NGC 6503 and UGC 5139) or there is not enough coverage in the 
	extinction map to effectively bin the galaxy into radial bins (UGC 4305). There are 
	published average metallicities from \cite{croxall09} for UGC 4305 and UGC 5139 and from \cite{tikhonov14} for NGC 
	6503. We plot the average dust--to--gas ratios calculated with $N_H = N_{HI} + 2N_{H_2}$
	against metallicity in the left plot in Figure \ref{fig:dgrmetal}. The scatter in Figure \ref{fig:gascorrs} indicates
	that the weighted average extinction likely corresponds to a slightly different location in the line of sight 
	in each individual pixel. Some pixels may inherently have more stars behind or in
	front of the dust layer. An average of the $A_V$ and $N_H$ will still be related to the average dust--to--gas ratio 
	modulo a factor two, for those galaxies where we can probe through the dust layer (see Table \ref{tab:extinctlimits} 
	estimates of maximum observable $A_V$). The factor might be larger than two for more dusty or more distant galaxies where we do not fully sample 
	the dust layer. We calculate average $A_V$ of all the pixels with at least 10 stars, and an 
	average $N_H$ of their correlated gas column densities, then calculate $A_V/N_H$ from these average values.
	
	There is a clear trend of decreasing dust--to--gas ratio with 
	decreasing metallicity. The lines in the plot are linear fits to the data on a logarithmic scale, implying a power law 
	relationship between metallicity and dust--to--gas ratio. The dashed line is the fit to the spiral galaxies
	only, while the solid line shows the fit to the entire sample of points. We see very little scatter in our 
	sample given the fairly high uncertainty in the literature metallicity	values. This could imply an 
	overestimation in these metallicity uncertainties. 
	
	We do not have access to CO data of a similar resolution to the HI data for NGC 7793, so its $N_H$ values are lower limits. 
	The dust--to--gas ratios for this galaxy are therefore upper limits. UGC 5139 and UGC 4305 also both have negligible H$_2$
	column densities compared to the HI. This may be more an effect of our choice of a constant $X_{CO}$ factor, so it's possible
	that the dust--to--gas ratios for these two galaxies are also upper limits.
	We therefore plot the dust--to--gas ratio for each galaxy calculated using the HI column density against the
	metallicity in the right plot in Figure \ref{fig:dgrmetal}. We still observe a correlation between dust--to--gas ratio and
	metallicity, however, as seen in Table \ref{tab:metalslopes}, the slope is steeper than in the full $N_H$ case. Clearly, 
	$N_{H_2}$ is important for determining the relationship between dust--to--gas ratio and 
	metallicity. If we underestimate $N_{H_2}$ in low-metallicity galaxies by using a constant $X_{CO}$, the 
	dust--to--gas ratios for these galaxies may in fact be upper limits, and the slope of the relation may be 
	steeper than what we observe. This is discussed further in section \ref{sec:dgrvmetal}.  
	
	Table \ref{tab:metalslopes} gives the slopes of the linear fit from Figure \ref{fig:dgrmetal}. \cite{remy14}
	give a slope of $1.6 \pm 0.3$ for a single power law, and use a slope of 1 as determined by previous studies
	\citep{james02, draine07, galliano08, leroy11} for the high metallicity end	($12+\log(O/H) > 7.96 \pm 0.47$). 
	All of our points lie within this metallicity regime, so it is difficult to determine if a broken power law
	is necessary to explain the changing dust--to--gas ratio with metallicity. 
	
%	\begin{figure}[h]
%        	\centering
%            \plottwo[width=0.5\textwidth]{allgals_dgrmetals_noCO_masscut14.pdf}{
%            \caption{Average $\frac{A_V}{N_{HI}}$ vs. metallicity for all NGC 628 radial bins, NGC 6503, NGC 7793 
%            (both radial bins), UGC 5139, and UGC 4305. Red points use the ``maximum" extinction
%            maps, while blue use the average extinction maps. The black line is the typically observed Milky 
%            Way value. Vertical errorbars are determined using the scatter of both $A_V$ and $N_{HI}$ in that bin 	
%            or galaxy. The black horizontal errorbar in the upper left shows the typical metallicity uncertainty
%            from the literature. Dashed lines are fit to NGC 6503 and the NGC 628 radial bins, while solid
%            lines are fit to the entire sample.}
%            \label{fig:dgrmetal_noCO}
%    \end{figure}
	
	\begin{deluxetable}{ccccc}
		\tablewidth{\textwidth}
		\tabletypesize{\footnotesize}
		\tablecolumns{5}
		\tablecaption{Dependence of the dust--to--gas-ratio on metallicity
			\label{tab:metalslopes}}
		\tablehead{
			\colhead{} & \multicolumn{2}{c}{$N_H$} & \multicolumn{2}{c}{$N_{HI}$ Only} \\
			\colhead{} & \colhead{Average} & \colhead{``Maximum"} & \colhead{Average} & \colhead{``Maximum"}
			}
		\startdata
			Spirals Only & 	$1.22 \pm 0.16$ &	$1.30 \pm 0.12$ & $2.39 \pm 0.17 $ & $2.47 \pm 0.18$ \\
			Full Sample	 &	$1.46 \pm 0.06$  &  $1.30 \pm 0.04$	& $1.82 \pm 0.07$  & $1.66 \pm 0.08$\\
		\enddata
		\tablecomments{All values are the slopes of the linear fits from Figure 	
			\ref{fig:dgrmetal}.}
	\end{deluxetable}
	
	\section{Discussion}
		\label{sec:discussion}	
		
		\subsection{Dust--to--Gas Ratio vs. Metallicity}
		\label{sec:dgrvmetal}		
		The slopes for the entire sample shown in Table \ref{tab:metalslopes} are consistent within the uncertainties 
		with the \cite{remy14} single power law slope for a fixed $X_{CO}$, despite being limited to half the metallicity 
		range and having a large gap between the spirals and the dwarf galaxies. The case where 
		we exclude CO and fit the spirals only is slightly steeper. This is more consistent with the single
		power law slope of $2.02 \pm 0.3$ from \cite{remy14} for a varying $X_{CO}$. Both cases are steeper than 
		the slope of 1 for the high-metallicity end of the broken power law relationship 
		in \cite{remy14}. Due to geometric effects, we tend to underestimate the extinction. This effect is 
		magnified in areas of high extinction. A correction for this effect might result in a steeper 
		slope than what we observe, as high-metallicity, highly-extinguished regions are pushed to higher extinctions 
		and higher dust-to-gas ratios faster than the low-metallicity, more lowly-extinguished regions. 
		This would not solve
		the discrepancy, so	we find dust--to--gas ratios that are inconsistent with a broken power law with a 
		high-metallicity end slope of 1. Our two dwarf galaxies, UGC 4305 and UGC 5139, are roughly at the break 
		in the power law from \cite{remy14}, so we cannot determine if we are consistent with the slope of the 
		low-metallicity end, or even if our data require a broken power law. With the full LEGUS sample, we can 
		better test if a varying $X_{CO}$ based on a galaxy's metallicity leads to a more consistent agreement 
		with previous work and reduced scatter in the relations between dust--to--gas ratio and metallicity.
		
		As discussed in Section \ref{sec:radial}, it is also likely that we underestimate
		the dust--to--gas ratio of the radial bins of NGC 628, especially in the center,
		since we are likely seeing a high percentage of blends. Highly obscured faint
		stars will be washed out by the less-obscured faint stars, reducing the 
		extinctions we derive for a region. NGC 628 is likely the furthest galaxy
		this method should be attempted on at the angular resolution of our survey.
		
		For this work, we have assumed the constant $X_{CO} = 2 \times 10^{20}$ cm\textsuperscript{2} K\textsuperscript{-1} 
	  	(km/s)\textsuperscript{-1} used in \cite{rahman11} and
		\cite{leroy08}, however there is evidence that $X_{CO}$ varies with metallicity \citep{leroy11, bolatto13} and
		star formation rate \citep{clarkglover15}. 
		Low-metallicity galaxies with a lower star formation rate like UGC 5139 and UGC 4305 likely have a higher $X_{CO}$ 
		than the spiral galaxies NGC 628 and NGC 6503, which would drive their gas content up, specifically the H$_2$ gas. 
		This has the overall effect of reducing the dust--to--gas ratios calculated for these galaxies. \cite{remy14} found 
		that an $X_{CO}$ with a metallicity dependence of $(O/H)^{-2}$ caused shifts in the dust--to--gas ratios of half
		a dex or less across the entire sample of galaxies, though localized dust--to--gas ratios will vary based
		on the relative column densities of HI and CO in the map. If this shift is dependent on metallicity, there could
		be a change in our observed slope of metallicty vs. dust--to--gas ratio. A higher $X_{CO}$ at lower metallicity 
		would increase the gas column density at low metallicity but not at high metallicity, increasing our observed
		slopes further. \cite{bolatto13} also find that 
		there is evidence that the $X_{CO}$ varies within galaxies, leading to a potential cause of scatter in our 
		gas correlation plots for each galaxy.	
		
		By requiring a minimum of 10 stars per pixel to calculate an extinction in 
		the extinction maps, we potentially introduce a bias towards lower extinctions.
		Regions with high extinction are likely to have few stars that are above our magnitude
		limits, which are indistinguishable from regions that simply do not have many stars
		in our maps. When we relax the requirement to a minimum of 4 stars per pixel and repeat the
		full analysis, there is a slight shift in the dust--to-gas ratios. However, the shift in the
		measured average extinction is less than the standard deviation across the map in each case. The 
		dust--to--gas ratios of galaxies shift a few tenths of a dex at most.
		A fit of the ratios derived from the weighted average maps for the 4 star case
		results in a slope of $1.09 \pm 0.06$ for the full sample when CO is included and $1.33 \pm 0.11$ 
		for the disks alone including CO. When CO is excluded, it results in $1.50 \pm 0.10$ for the full sample  
		and $2.05 \pm 0.15$ for the disks alone.
		
		Our choice to only select stars more massive than 14 M$_\odot$ could also result in an overestimation of the dust--to--gas ratio. Dust is likely concentrated around star-forming regions where young, massive stars are located. \cite{zaritsky02} found a 0.3 magnitude increase in $A_V$ for young massive stars as compared to older, less massive stars in the SMC. If this increase is constant across all galaxies, it is possible
		that our dust--to--gas ratios are overestimates, but should not cause a change in our derived slope
		for the relation between dust--to--gas ratio and metallicity.		

		\subsection{Comparison to Extinctions Derived from Spectral Energy Fitting}
		\label{sec:beastcomp}
		The Bayesian Extinction and Stellar Tool (BEAST) is an SED-fitting method used by the PHAT survey to 
		calculate individual stellar extinctions in M31 using multi-band photometry. \cite{gordon16} 
		have recently made this software publicly available. It serves as a useful comparison to our individual 
		extinctions calculated from the isochrone-matching method described in Section \ref{sec:extinctcorrs} for
		massive stars, since both methods use the same input data and work with individual massive stars. Other methods of extinction corrections (i.e., from IR emission maps) might not specifically 
		probe the massive stars or might probe different regions of gas and dust. A comparison with BEAST allows
		a star-by-star comparison between methods.
		BEAST allows the simultaneous fitting of several different extinction parameters, including $A_V$ and $R_V$. 
		Here we compare the BEAST derived extinctions to those of the 
		isochrone-matched extinctions from Section \ref{sec:extinctcorrs} for the eastern field of NGC 7793. We
		employ identical magnitude- and Q-based cuts to both the isochrone-matched stars and the BEAST derived stars.
		This is the closest spiral in our sample, so should be least affected by crowding and the extinction law should be 
		fairly MW-similar (i.e., not starburst-type, as in \cite{calzetti94}). For our comparison, we restricted the 
		BEAST extinction law to a MW-type extinction law with $R_V = 3.1$, which corresponds to the reddening slope 
		used in Section \ref{sec:extinctcorrs}. The SED-fitting was only performed on stars with robust PSF fits in all 
		five bands with photometric errors $\leq$ 0.25 magnitude.
		
\begin{figure}[h!]
        	\centering
        	\epsscale{2.3}
            \plottwo{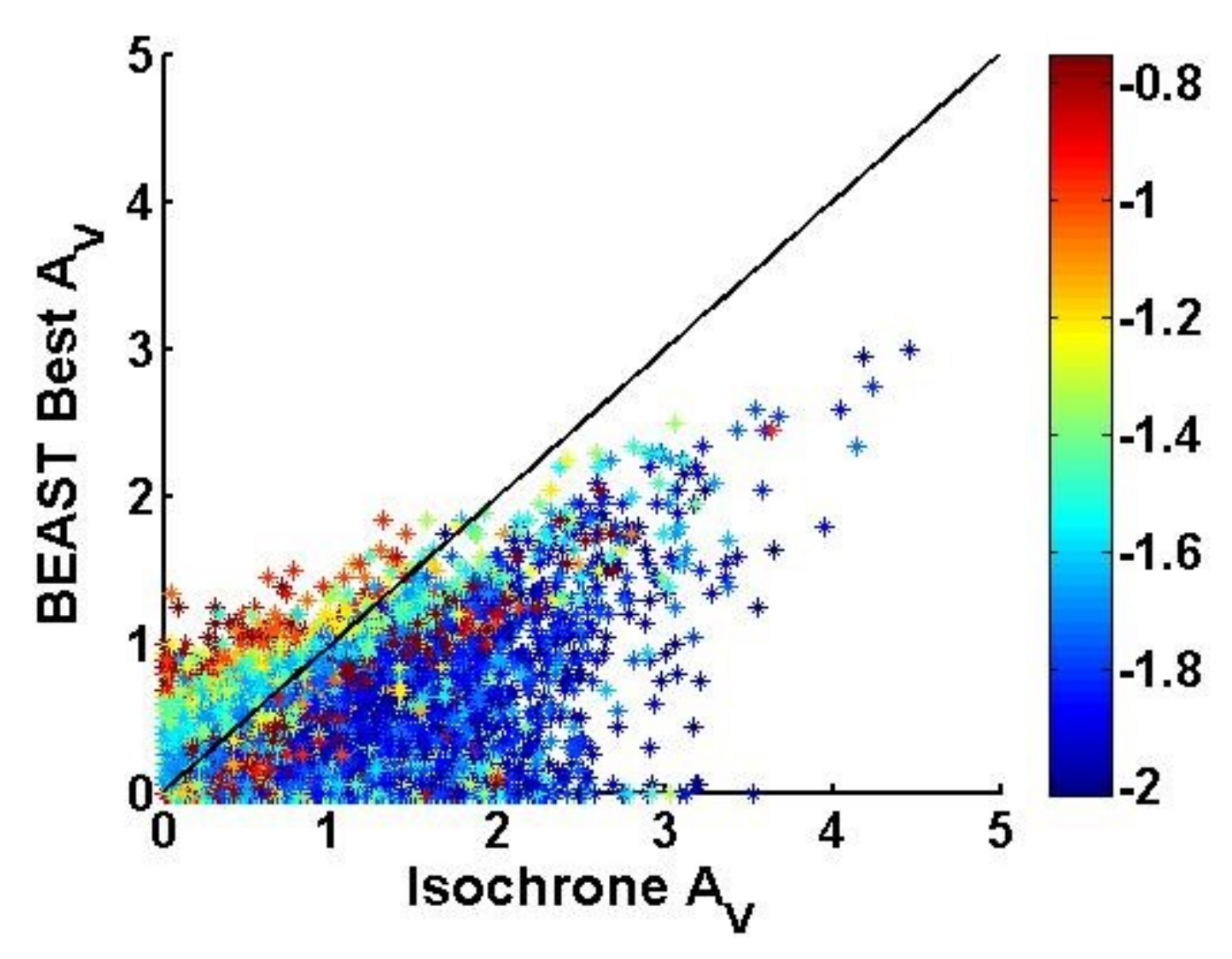}{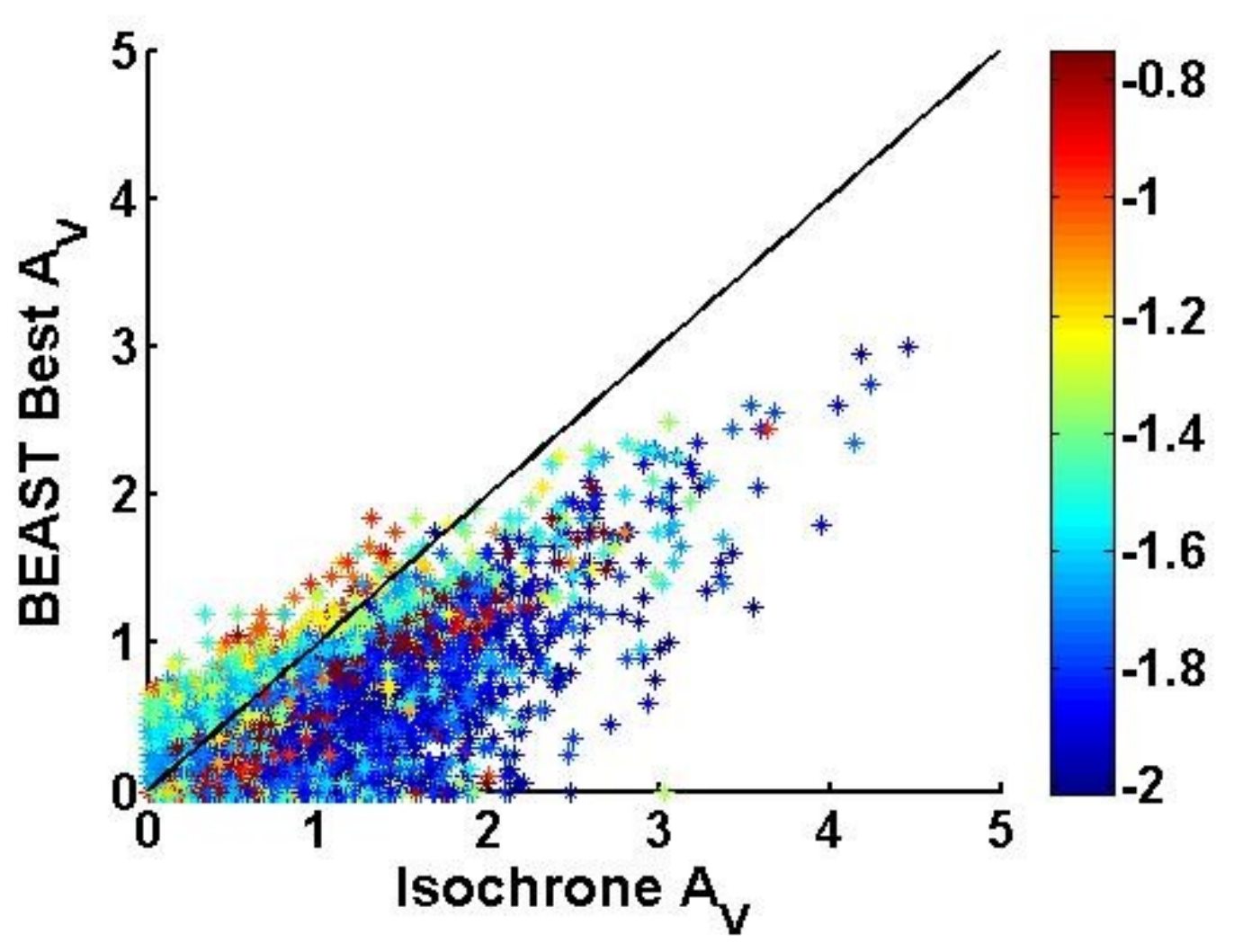}
            \caption{Comparison of $A_V$ calculated from isochrone matching and BEAST SED-fitting. Isochrone-
            matched extinctions are on the x-axis, BEAST extinctions on the y-axis. Points are colored with
            respect to $Q_{NBVI}$, with more negative values corresponding to the brightest, most massive stars.
            The black solid line shows the one-to-one values, while the black dashed line shows the linear fit
            to all points, with the equation in the top left. Top: All corrected stars. Bottom:
            all corrected stars with an F275W absolute magnitude brighter than or equal to -5 as 
            determined by the isochrone-matching method.}
            \label{fig:beastcompQ}
    	\end{figure}

		Figure \ref{fig:beastcompQ} shows a comparison between the calculated $A_V$ for individual stars
		for our isochrone-matching method and the BEAST SED-fitting method, color-coded	by $Q_{NBVI}$, for both the 
		entire corrected sample of BEAST-fitted stars and for BEAST-fitted stars above the absolute
		magnitude cut used in Section \ref{sec:extinctmaps} to generate the extinction
		maps. 
		
\begin{figure}[h!]
        	\centering
        	\epsscale{2.3}
           \plottwo{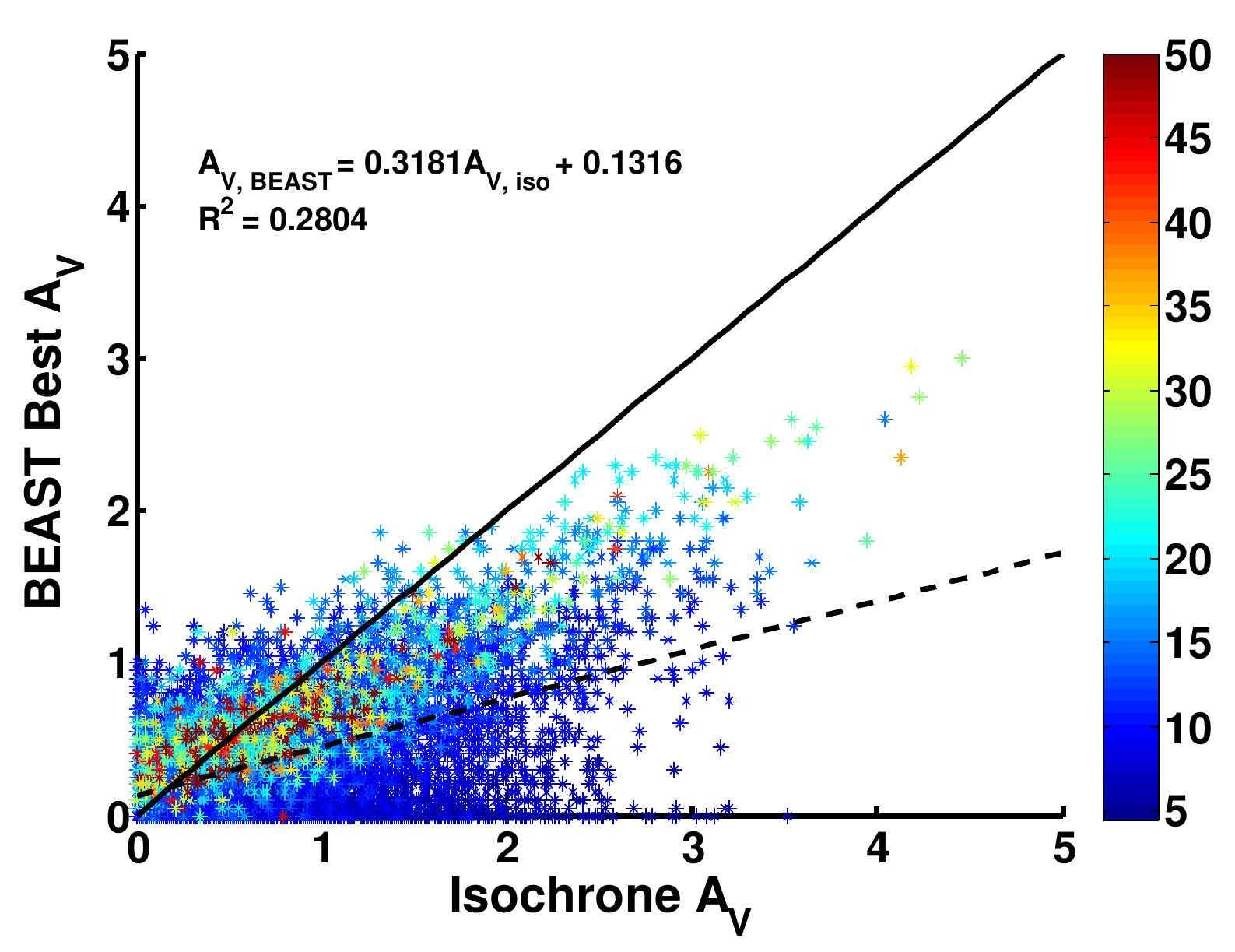}{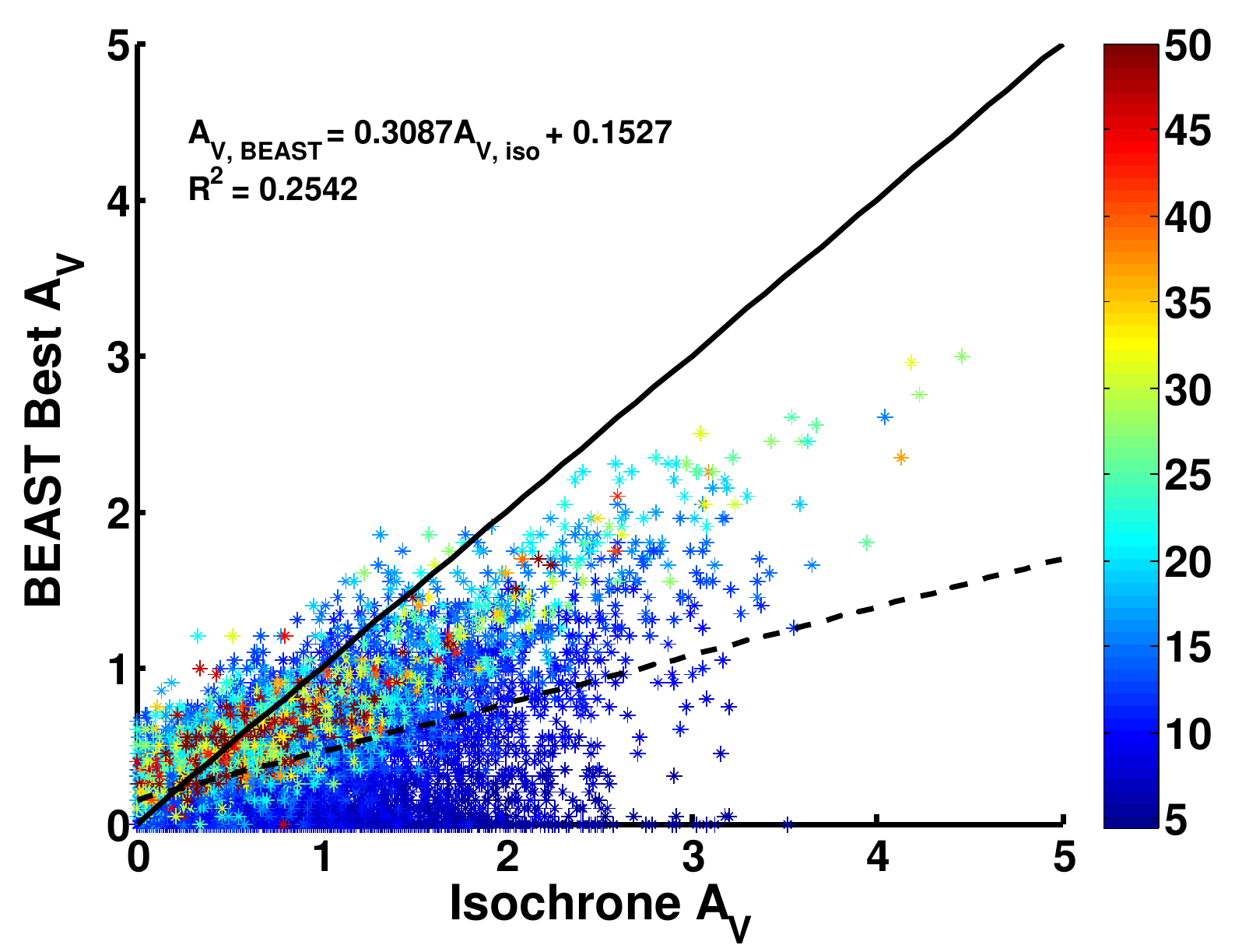}
            \caption{Same as figure \ref{fig:beastcompQ}, but with points colored according to BEAST-derived
            stellar mass.}
            \label{fig:beastcompmass}
    	\end{figure}

\begin{figure}[h!]
        	\centering
        	\epsscale{2.3}
           \plottwo{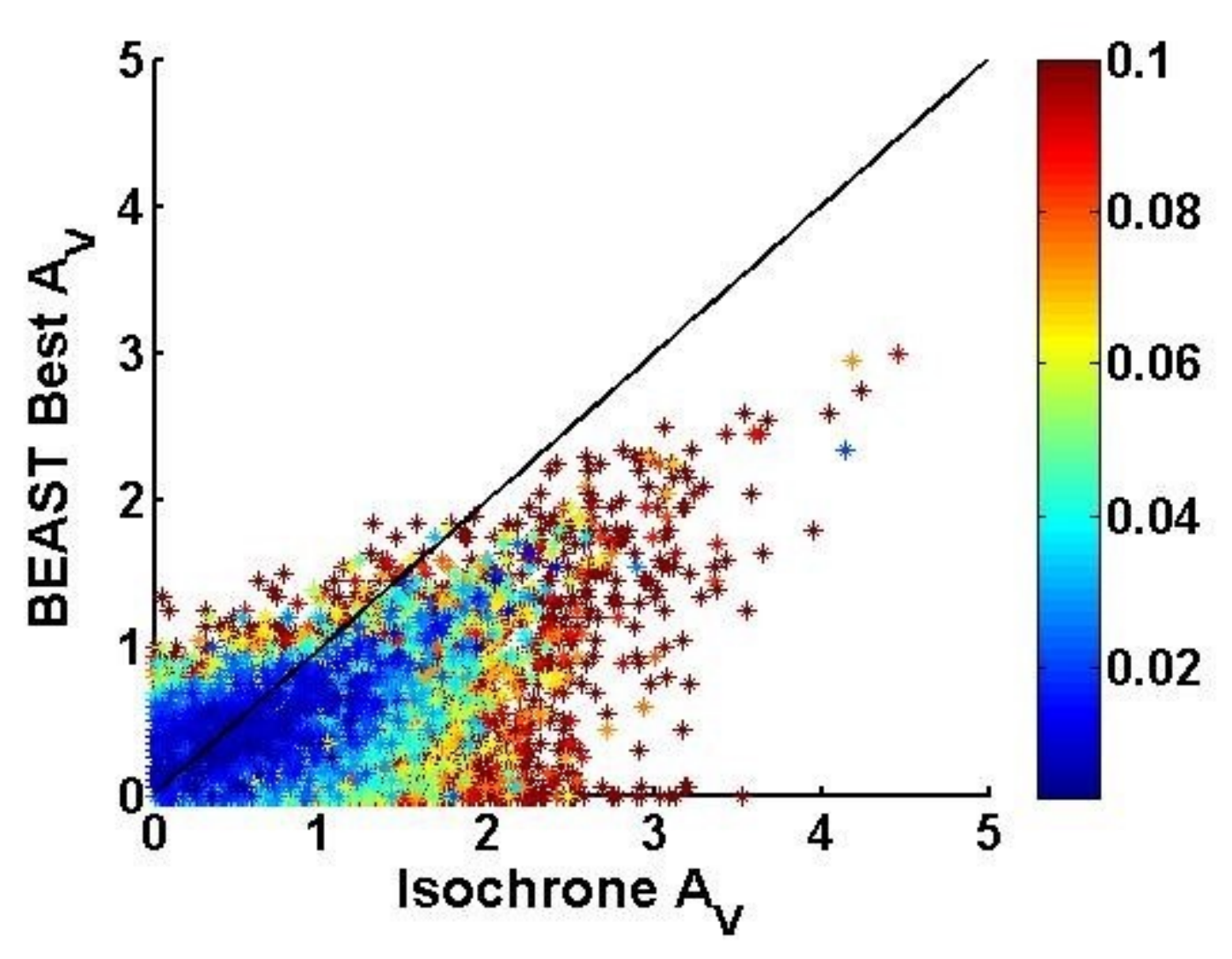}{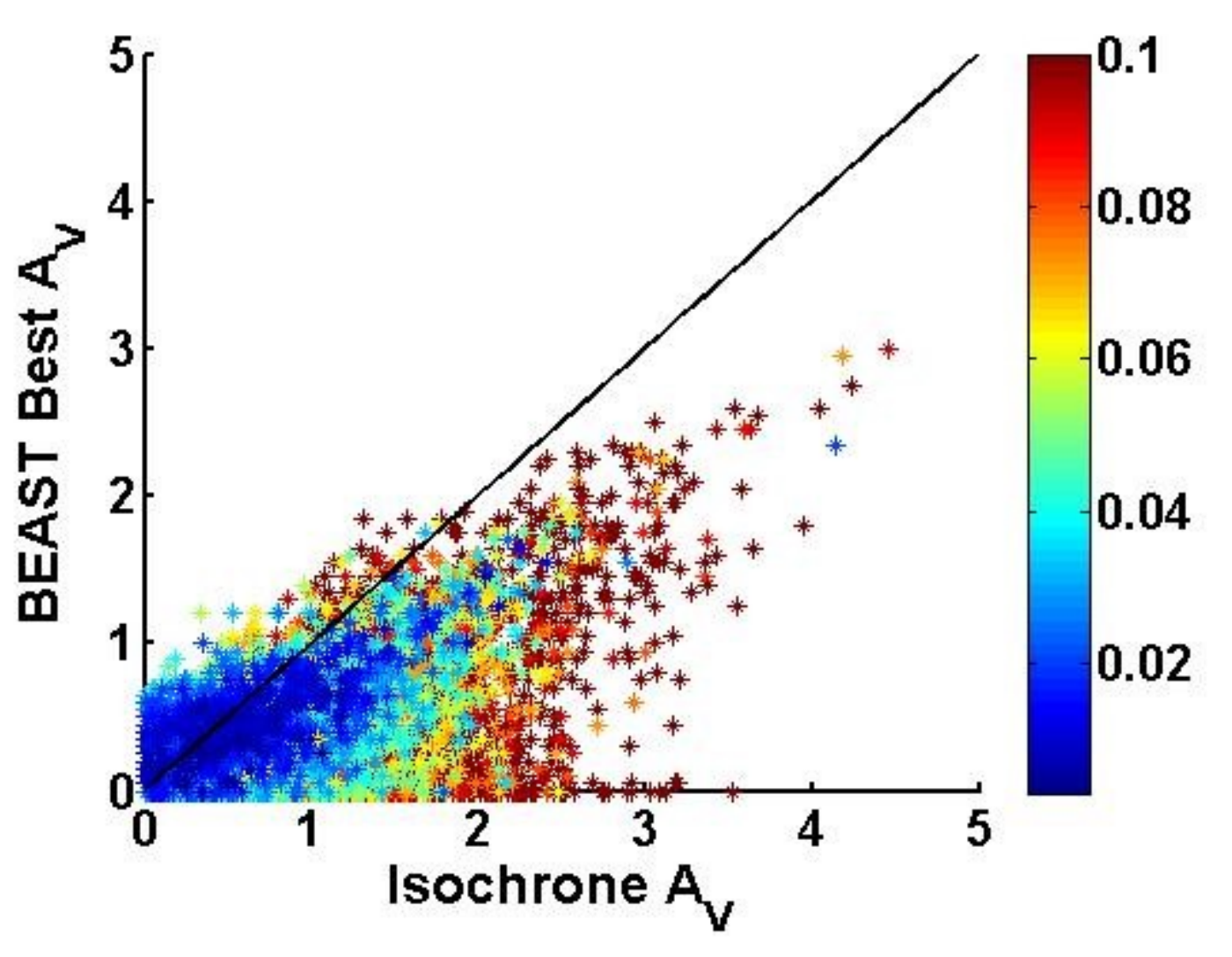}
            \caption{Same as figure \ref{fig:beastcompQ}, but with points colored according to the observed 
            photometric error in F275W.}
            \label{fig:beastcomperr}
    	\end{figure}

\begin{figure}[h!]
        	\centering
        	\epsscale{2.3}
           \plottwo{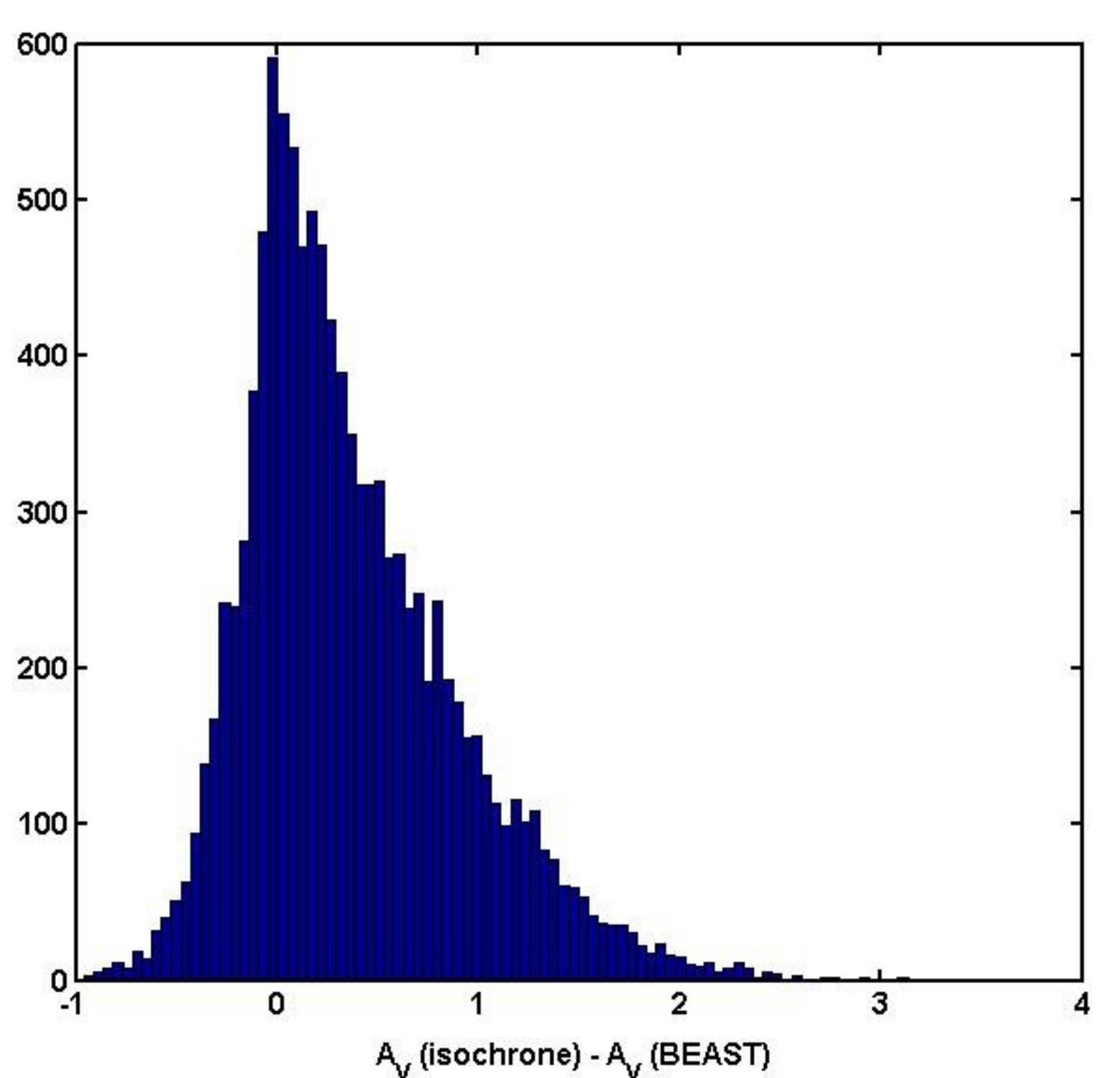}{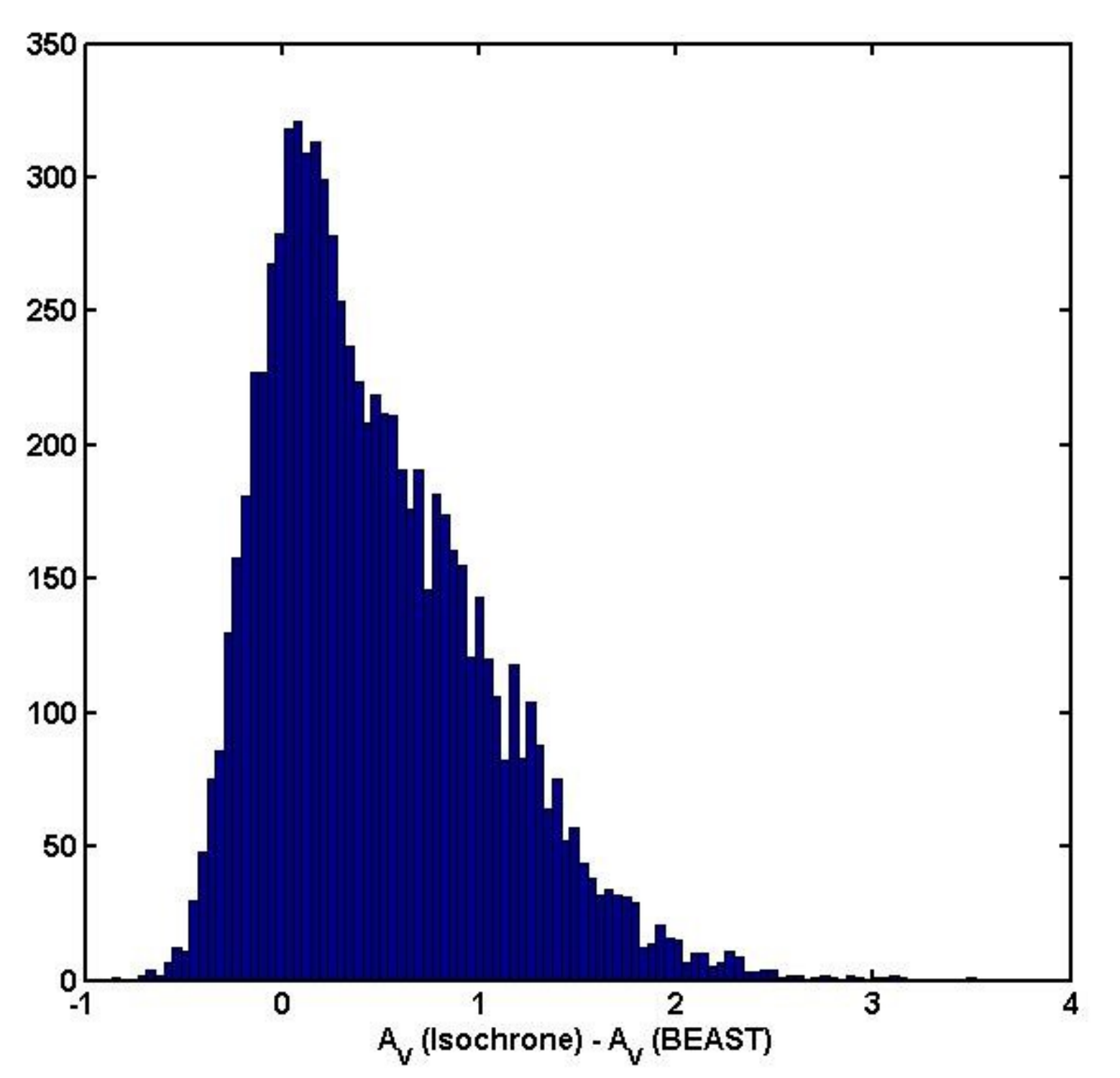}
            \caption{Histogram of $A_V(isochrone) - A_V(BEAST)$ for individual stars in the eastern field of
            NGC 7793. Top: All corrected stars. Bottom: All corrected stars with an F275W absolute magnitude
            brighter than or equal to -5 as determined by the isochrone-matching method.}
            \label{fig:diffhists}
    	\end{figure}

		BEAST tends to assign systematically lower $A_V$ to stars compared to the isochrone method, especially at 
		higher extinctions. There is also a large population of stars that BEAST assigns
		low mass, but the photometric data show a $Q_{NBVI}$ and two-color values that place the stars in
  		the range of massive stars, as seen in
		Figures \ref{fig:beastcompQ} and \ref{fig:beastcompmass}. Since there are generally many more low-mass stars than high mass stars,
		it is possible that BEAST automatically assumes, given photometric equivalence, that the star is low-mass
		as opposed to high mass. After applying the absolute magnitude cut
		used in Section \ref{sec:extinctmaps}, many of the stars in this grouping are
		removed. Figure \ref{fig:beastcomperr} also shows that
		the stars with the greatest differences tend to be the stars with the highest
		photometric errors, especially in the population of stars with low assigned 
		BEAST masses and $Q_{NBVI}$ values corresponding to high masses. The systematic difference between derived
		extinctions using BEAST and isochrone matching is even more obvious in Figure \ref{fig:diffhists}, which shows
		the distribution of $A_{V}(isochrone) - A_{V} (BEAST)$ for individual stars. The peak of the distribution is
		positive, with a large tail further to the positive.		
		
    	An additional source of uncertainty is the effect of post-MS stars, which can vary in color by as much as 0.1. This
    	is within our photometric error, so should not be much of an effect, but if BEAST puts more weight on the post-MS,
    	then it could account for at least some of the systematic difference.
    	
    	The average extinction map generated using the BEAST extinctions reflects that the BEAST individual stellar 
    	extinctions tend to be systematically lower than the isochrone-matched extinctions. This 
    	can lead to differences in the extinctions of about 50\% across the map. We have thus far only tested 
    	a single field of one galaxy, and do not yet know if this will affect the relative changes in the dust--to--gas ratio across galaxies. More study is needed to examine
    	the exact nature of the differences in extinction and if this behavior is consistent across other 
    	galaxies, specifically dwarf galaxies where we already find very low extinctions. 
		
    \section{Summary and Future Work}
        \label{sec:summary}   
        
        We correct the photometry of massive individual stars for extinction using the isochrone matching method used by
        \cite{kim12}, and compare these extinction corrections to those derived from SED--fitting for a single pointing. 
        We generate extinction maps using the extinction values of individual massive stars, averaged over 
        spatial bins to account for the range of optical depths along the line-of-sight. This is the first time that
        such maps have been derived using massive stars for galaxies outside the Local Group, and this method can be expanded to other galaxies within
        10 Mpc. Extinction maps such as these may be useful in probing cold and dark gas and dust that may be missed in 
        emission \citep{paradis12}. We correlate the extinction maps with neutral gas maps in order to investigate the dust--to--gas mass ratio
        in NGC 628, NGC 7793, NGC 6503, UGC 4305, and UGC 5139. Our results can be summarized as follows:
        
        \begin{itemize}
        	\item We find lower extinctions across the dwarf galaxies UGC 4305 and UGC 5139 than in the spiral galaxies
        	NGC 628, NGC 7793, and NGC 6503.
        	\item There is wide scatter in localized dust--to--gas ratios for each galaxy. This could be explained by geometric
        	effects caused by variations in the position of stars along the line--of--sight.
        	\item NGC 628 shows a significant radial variation in dust--to--gas ratio with galactocentric radius, likely 
        	correlated with its metallicity gradient.
        	\item We find strong evidence to support a power law relationship between dust--to--gas ratio and metallicity consistent with other
        	work. A metallicity-dependent CO-to-H$_2$ conversion factor $X_{CO}$ would increase the slope compared to what is observed here.
        	\item There is a systematic difference in the extinctions derived via isochrone matching and those derived using 
        	the BEAST SED--fitting algorithm. The cause of this systematic difference may be related to a preference within 
        	BEAST to assign low mass to stars for which the Q-value would suggest a high mass.
        \end{itemize} 
        
        Further investigations will provide a better understanding of the relationship between dust--to--gas
        ratio and metallicity, as a larger sample of LEGUS galaxies will expand and fill in our metallicity 
        range. We will expand our comparison of the SED-fitting derived extinctions with isochrone-matching
        derived extinctions. We will investigate the possiblity of a relationship between dust--to--gas ratio and 
        galaxy morphology and star formation rate. If either of these factors influence dust--to--gas ratio,
        it could explain the scatter seen in the dust--to--gas mass ratio seen across our extinction maps, since the
        uncertainties in our individual pixels can only account for $\sigma_{A_V} \sim 0.1$.
                
        Our 1"x1" gridded adaptive resolution maps have a variety of applications. They can be used to 
        correct H$\alpha$ and other optical images for extinction, including correcting stars in the LEGUS catalogs 
        that cannot be corrected with isochrone matching. These stars may have been below our magnitude or Q-value
        cutoffs, but if they are located near stars corrected with isochrone matching, they likely have similar
        extinctions, allowing us to apply the extinctions from the map. Corrections to H$\alpha$ maps are useful
        for determining accurate star formation rates. It is also possible to match these maps to maps of the 
        diffuse ionized gas (DIG) and HII regions in order to study differences in extinction between these two
        important phases of the ionized ISM. This is important for determining the fractional contribution of
        the DIG to the total H$\alpha$ luminosity, which in turn can help determine ionization sources for the 
        DIG \citep{walterbos94}.
        
        This work is based on observations made with the NASA/ESA Hubble Space Telescope, obtained at the Space Telescope
        Science Institute, which is operated by the Association of Universities for Research in Astronomy, under
        NASA Contract NAS 5-26555. These observations are associated with Program 13364 (LEGUS). Support for 
        Program 13364 was provided by NASA through a grant from the Space Telescope 
        Institute. L. K. and R. A. W. acknowledge support from the Space Telescope Science institute 
        [Grant Numbers HST-GO-13773.003-A]. 
        L. K. would additionally like to acknowledge support from Gemini Observatory in La Serena, Chile. 
        M. F. acknowledges support from the Science and Technology Facilities 
        Council [Grant Number ST/P000541/1]. D. A. G. kindly acknowledges financial support 
        by the German Research Foundation (DFG) through programme GO 1659/3-2. G. A. acknowledges support from the 
        Science and Technology Facilities Council (ST/P000541/1 and ST/M503472/1). We would also like to thank the
        anonymous referee for their useful comments which helped improve the paper.
        
        \software{BEAST \citep{gordon16}, DOLPHOT photometric package \citep{dolphin02}, PARSEC \citep{parsec}}

\bibliographystyle{apj}
\bibliography{research}

\end{document}